\documentclass[12pt,a4paper,english]{article}
\usepackage[T1]{fontenc}
\usepackage[latin2]{inputenc}
\usepackage[english]{babel}
\usepackage{amsmath}
\usepackage{amsfonts}
\usepackage{indentfirst}
\usepackage[dvips]{graphicx}

\newcommand{\bea}{\begin{eqnarray}}
\newcommand{\eea}{\end{eqnarray}}
\newcommand{\be}{\begin{equation}}
\newcommand{\ee}{\end{equation}}

\newcommand{\hf}{\frac{1}{2}}

\newcommand{\cZ}{{\mathcal{Z}}}
\newcommand{\cF}{{\mathcal{F}}}

\newcommand{\Li}{{\rm Li}}

\def\Tr{{\rm Tr \,}}

\def\G{\Gamma}

\begin{document}

\sloppy

% TITLE PAGE

\begin{flushright}
\begin{tabular}{l}
BONN-TH-2008-07

\\ [.3in]
\end{tabular}
\end{flushright}

\begin{center}
\Large{ \bf Seiberg-Witten theory and matrix models}
\end{center}

\begin{center}

\bigskip

Albrecht Klemm$^1$ and Piotr Su{\l}kowski$^{1,2}$

\bigskip

\medskip

\emph{$^1$ Physikalisches Institut der Universit{\"a}t Bonn and Bethe Center for
  Theoretical Physics,} %\\
\emph{Nussallee 12, 53115 Bonn, Germany} \\  [4mm]
\emph{$^2$ So{\l}tan Institute for Nuclear Studies,} %\\
\emph{ul. Ho\.za 69, 00-681 Warsaw, Poland} \\ [4mm]

\bigskip   

\centerline{\emph{aklemm@th.physik.uni-bonn.de, Piotr.Sulkowski@fuw.edu.pl} }

\smallskip
 \vskip .6in \centerline{\bf Abstract}
\smallskip

\end{center}

We derive a family of matrix models which encode solutions to the Seiberg-Witten
theory in 4 and 5 dimensions. Partition functions of these matrix models 
are equal to the corresponding Nekrasov partition functions, and their
spectral curves are the Seiberg-Witten curves of the 
corresponding theories. In consequence of the geometric engineering,
the 5-dimensional case provides a novel matrix model formulation of the topological string theory
on a wide class of non-compact toric Calabi-Yau manifolds.
This approach also unifies and generalizes other matrix models, such as
the Eguchi-Yang matrix model, matrix models for bundles over $\mathbb{P}^1$, and Chern-Simons matrix 
models for lens spaces, which arise as various limits of our general result.

%*******************************************************
%*******************************************************

\newpage

\tableofcontents

\newpage

%*******************************************************

\section{Introduction}

Finding the solution of $\mathcal{N}=2$ supersymmetric gauge theories by
Seiberg and Witten in terms of associated 
families of hyperelliptic Riemann surfaces, the Seiberg-Witten curves \cite{SW-1,SW-2}, was a significant  
development in theoretical physics. The so-called Seiberg-Witten theory unifies many branches of physics and mathematics. 
From the physics perspective generalizations to theories with various gauge groups and matter contents were studied, 
as well as their relation to and the embedding in string theory. From the mathematical viewpoint the Seiberg-Witten 
solution gave important insights into the topology of four-manifolds. Also the solubility of $\mathcal{N}=2$ 
theories turned out to be related to the underlying integrability, and the relations to various integrable systems 
were found. The literature on all these developments is immense, and the good starting point might be to consult 
the following reviews \cite{bilal-sw,klemm-sw,dhok-phong-sw,cal-inst,iga-sw,lm}, as well as references therein. 

The holomorphic partition function $\cZ$ is a central object in the theory. Its 
asymptotic expansion $\cZ=\exp \sum_{g=0} \hbar^{2g-2} \cF_g({\bf a},\Lambda)$ in $\hbar$ is 
a generating function for gauge theory instanton numbers, which appear as 
coefficients of $\hbar$, and inverse powers of the vevs $a_l$ in the
diagonal of the adjoint Higgs field. In $a_l^{-1}$ the $\cF_g({\bf
  a},\Lambda)$ have a finite radius of convergence as it has to be the case 
for physical terms in the effective action.  In particular 
$\cF=\cF_0({\bf a},\Lambda)$ is the prepotential, which determines the exact 
low energy gauge theory effective action up to two derivatives. The
$\cF_g({\bf a},\Lambda)$ for  $g>0$ multiply gravitational couplings of the form $R_+^2
F_+^{2g-2}$, where $R_+$ and $F_+$ are the self-dual parts of the curvature and
the graviphoton field strength respectively. 

The fact that $\cF_0({\bf a},\Lambda)$ can be calculated from periods of  
a family of hyperelliptic Riemann surface $\Sigma({\bf a},\Lambda)$ 
over a meromorphic one form differential $dS$  was the main insight of \cite{SW-1,SW-2}. 
However this was argued using global  consistency conditions of the low energy
effective action and not from the microscopic action itself. The latter
argument was provided by Nekrasov and Okounkov~\cite{Nek,Nek-Ok}. 
In~\cite{Nek} Nekrasov developed a direct instanton calculcus and used 
localization techniques to determine $\cZ$ to all orders in $\hbar$ as sums over 
two-dimensional partitions labeling the instanton configurations. 
In the thermodynamic limit the limiting shape of the partitions approaches 
the Seiberg-Witten curve $\Sigma({\bf a},\Lambda)$ and in~\cite{Nek-Ok} it was 
proven that the periods of the latter over $dS$  reproduce $\cF$ to all 
orders in $a_l$. 

The curve $\Sigma({\bf a},\Lambda)$ and the differential $dS$ are the defining data of the 
$\mathcal{N}=2$ supersymmetric theories in the sense that  $\cZ$ can be 
reconstructed from them. In particular one can view 
$\Sigma({\bf a},\Lambda)$ as the spectral curve of a 
putative matrix model and the A-cycle integrals of $dS$ 
as the fixed filling fractions and  use the recursive 
solutions~\cite{eyn-or} of the loop equation to 
reconstruct the $\cF_g$. 

In this paper we find the explicit 1-matrix model for $SU(n)$ Seiberg-Witten
theory, whose spectral curve is the Seiberg-Witten curve. The strategy we use is to 
represent partitions by $N\times N$  matrix integrals in a way recently proposed by Eynard~\cite{eynard-planch}, 
which explicitly reproduces the Nekrasov partition function in the 
large $N$ limit. The matrix model which we find reads
\be
Z^{4d} =  \int_{Mat_{nN}} \mathcal{D}M e^{-\frac{1}{\hbar} \Tr V^{4d}(M)},   \qquad \qquad \mathcal{D}M=\prod_i dx_i \prod_{i<j}(x_i-x_j)^2, 
\ee
where $M\in Mat_{nN}$ is $nN\times nN$ matrix and the measure $\mathcal{D}M$ involves the ordinary Vandermonde determinant. In the large $N$ limit the potential is given by
$$
V^{4d}(x) = tx + 2\sum_{l=1}^n \Big( (x-a_l)\log(x-a_l) - (x-a_l) \Big),
$$
where $a_l$ are vevs of the Higgs field, while $t$ in the linear term 
encodes in particular the scale $\Lambda$. 

More precisely the fact that it is possible to write the Nekrasov partition function as a 
matrix integral is a consequence of its close relation to the Plancherel 
measure on partitions. The Nekrasov sum for $SU(n)$ theory can be 
regarded as a generalization of the Plancherel measure to the case of $n$ 
sets of partitions; equivalently it can be written as sum over one set of 
the so-called blended partitions. The properties of the Plancherel measure 
have been known for a long time; in particular it is known that in the thermodynamic limit it leads to the 
smooth limiting shape of large partitions known as the arc-sin
law~\cite{vershik-kerov,logan-shepp}. This was  used by \cite{Nek-Ok} to show
that analogous limiting shapes arise for the case of the Nekrasov 
partition function, and that they encode the geometry of Seiberg-Witten curves
of the corresponding $\mathcal{N}=2$ theories. Then, recently Eynard
demonstrated how to rewrite the ordinary Plancherel measure, as 
well as its $q$-deformation, as a matrix integral \cite{eynard-planch}. 
In this paper we use similar methods to derive the above matrix models for $SU(n)$ theory. 

We also derive a matrix model for a 5-dimensional theory compactified on a circle. We find that it it reads
\be
Z^{5d} = \int_{Mat_{nN}}  \mathcal{D}M e^{-\frac{1}{g_s} \Tr V^{5d}(M)},\qquad\qquad  \mathcal{D}M = \prod_i d u_i  \prod_{i<j} \Big(2 \sinh \frac{u_i - u_j}{2}\Big)^2,
\ee
which is related to a deformation of the 4-dimensional result.
Now the measure is given by the deformed Vandermonde determinant, while the potential reads
\be
V^{5d}(u) = t u + \frac{n}{2} u^2 + 2\sum_{l=1}^n \Li_2(e^{u+a_l}). \label{5d-MM}
\ee
As is well known, e.g. from the geometric engineering of gauge theories \cite{geom-eng}, the partition function for 
5-dimensional theory is equal to the partition function of topological strings on appropriate geometry. 
Therefore our matrix models also provide a new formulation of topological string theory, in the large 
radius limit of geometries which admit a limit to $SU(n)$ theories. In particular, in the text we will provide matrix model expressions for various non-compact toric Calabi-Yau spaces.

A very important feature of the matrix models which we derive is the fact that
in various limits they reduce to other well-known matrix models. In
particular, the 4-dimensional model (which itself is a limiting case of the
5-dimensional model) is a direct generalization of the
Eguchi-Yang matrix model \cite{E-Y,E-H-Y} and reduces to it for $n=1$. The 5-dimensional model,
also for $n=1$, leads immediately to matrix models for line bundles over
$\mathbb{P}^1$ \cite{CGMPS}.
On the other hand, in the so called orbifold limit which involves $t=0$ and
suppression of $\Li_2$ terms, the 5-dimensional model becomes just the quadratic
matrix model with the deformed potential, which is the Chern-Simons matrix
model for lens spaces postulated in \cite{mm-lens,lens-matrix} and analyzed in \cite{hal-yas,hal-ok-yas}.
These other seemingly unrelated matrix models turn out to be just various corners of
a single matrix model landscape.

The paper is organized as follows. In section \ref{sec-SW} we recall a necessary background on Seiberg-Witten 
theory and the Nekrasov partition function. In section \ref{sec-4d} we derive a matrix model for 4-dimensional 
$\mathcal{N}=2$ $SU(n)$ theory and show that its spectral curve coincides with the Seiberg-Witten curve of 
$SU(n)$ gauge theory. In section \ref{sec-5d} a matrix model for 5-dimensional
theory and its spectral curve are derived and analyzed. In section   
\ref{sec-top-string} we discuss its relation to topological string theory. In
section \ref{sec-models} we recover other well-known matrix models as certain limits of
our most general matrix model. Section \ref{sec-disc} contains a discussion.

%*******************************************************
%*******************************************************

\section{Seiberg-Witten theory}   \label{sec-SW}

In this section we review the solution of the Seiberg-Witten theory in terms
of two dimensional partitions, as well as the underlying family of curves, and
set up a necessary notation. After introducing two-dimensional partitions and
the Plancherel measure, we discuss how its generalization leads to the
Nekrasov-Okounkov partition function in 4 and then in 5 dimensions.

\subsection{Partitions and Plancherel measure}

A partition $\lambda$ of an integer number $|\lambda|$ is a set of non-negative, non-increasing integers
$$
\lambda=(\lambda_1,\lambda_2,\ldots), \qquad \lambda_1\geq \lambda_2 \geq \ldots \geq 0, \qquad 
\sum_{i}\lambda_i = |\lambda|.
$$
A partition can also be presented as a Young tableaux given by the set of
aligned horizontal rows, each containing $\lambda_i$ boxes. The number of rows
of a partition is defined as the number of non-zero $\lambda_i$.
When considering statistical ensembles of partitions one can choose various
measures. A very important one is the Plancherel measure defined as
\bea
P(\lambda) =\displaystyle{
\left(\frac{{\rm dim} \lambda}{|\lambda|!}\right)^2}=&\displaystyle{\frac{1}{\left(\prod_{i=1}^k(\lambda_i+N-i)!\right)^2}\prod^N_{i<j}\left(\lambda_i-\lambda_j-i+j\right)^2} =  \nonumber \\
=&\displaystyle{   \prod^\infty_{i<j}
\Big(\frac{\lambda_i-\lambda_j-i+j}{j-i}\Big)^2=\prod_{\square\in \lambda} \frac{1}{hook(\square)^2}},   \label{planch}
\eea
where $hook(\square)$ is the Hook-length of a given box in $\lambda$,
$N$ is the number of rows, and ${\rm
  dim}\lambda$ is the dimension of the represenetation of the symmetric group
corresponding to the Young-Tableaux. In the thermodynamic limit with respect to 
the Plancherel measure the partitions approach a limiting shape given by the so-called arc-sin law 
\cite{vershik-kerov,logan-shepp}. This shape is described by 
the arcsin function, hence its name. An important 
fact is that, after the rescaling leading to the smooth limiting shape, the limiting partition has a finite 
length. This phenomenon is called the arctic circle property.

One can also consider a $q$-deformation of the Plancherel measure, denoted $P_q$, which arises by replacing the integers in 
(\ref{planch}) by its $q$-deformation
$$
[ h ] = q^{-h/2} -q^{h/2},  \qquad \qquad [ h ]!  =  \prod_{i=1}^h (q^{-i/2} - q^{i/2}),  
$$
so that
\be
P_q(\lambda) = \prod_{i<j}
\Big(\frac{[\lambda_i-\lambda_j-i+j]}{[j-i]}\Big)^2.   \label{planch-q}
\ee

\subsection{The Nekrasov partition function}

The partition function for $\mathcal{N}=2$, $SU(n)$ theory was derived by Nekrasov in \cite{Nek} and analyzed 
further in great detail in \cite{Nek-Ok}. It is a generalization of the Plancherel measure to the case of $n$ sets 
of partitions. It therefore can be written down as a sum over a set of $n$ partitions $\lambda^{(l)}_i$, with 
$l=1,\ldots,n$ labeling various partitions and index $i\geq 0$ denoting $i$'th row of a  given partition. This 
partition function depends on the scalar vevs $a_l$
%\be
%a_l=\hbar p_l   \label{a-p}, \qquad \quad \textrm{such that}\  \sum_l p_l = 0, 
%\ee
via $a_{lk}=a_l-a_k$ and it reads  \cite{Nek,Nek-Ok,NY-I,NY-II}
\bea
Z^{SU(n)} & = & \sum_{\vec{\lambda}=(\lambda^{(1)},\ldots,\lambda^{(n)})} \Lambda^{2n |\vec{\lambda}|} 
Z_{\vec{\lambda}}, \label{Nekrasov-PF}  \\
Z_{\vec{\lambda}} & = &
\prod_{(l,i)\neq (k,j)} \frac{a_{lk}+\hbar(\lambda^{(l)}_i - \lambda^{(k)}_j + j - i)}{a_{lk}+\hbar(j-i)}, \nonumber
\eea
where $|\vec{\lambda}|=\sum_{l=1}^n |\lambda^{(l)}|$. The sums in the above expression are performed over all 
possible partitions. To convert this partition function into a matrix model expression, below we consider 
the sums over partitions with at most $N$ non-zero rows. In this 
case Nekrasov partition function is recovered in large $N$ limit, where $N\to \infty$ and 't Hooft coupling $t=\hbar N$ is fixed. More generally, we could introduce an 
independent number of rows $N_l$ for each partition $\lambda^{(l)}$. However it would not affect the answer, because the final result is independent of $N$.

%*******************************************************
%*******************************************************

\subsection{Seiberg-Witten curve} \label{swcurve}

The solution of the Seiberg-Witten theory is encoded in the Seiberg-Witten curve. For $SU(n)$ gauge theory this 
is a hyperelliptic curve of genus $n-1$, which can be defined by the equation 
\begin{equation} 
\Lambda^n\Big(w+\frac{1}{w}\Big) = P_n(x),\qquad P_n(x) = x^n + u_1 x^{n-1} + \ldots + u_n, 
\label{sw}
\end{equation} 
for $w,x\in\mathbb{C}$, which is related to the familiar hyperelliptic form $y^2=P_n^2-4 \Lambda^{2n}$ 
by the identification $w=\frac{1}{2 \Lambda^n}(y+P_n)$. The A- and B-periods of the differential
$$
dS = \frac{1}{2\pi i} x\frac{dw}{w}
$$
of this curve correspond respectively to the vevs $a_l$ and derivatives of the prepotential $\frac{1}{2\pi 
i}\frac{\partial \mathcal{F}_0}{\partial a_l}$. 

Let us recall now how this curve emerges as the stationary point of functionals \cite{Nek-Ok}, which arise from the
ADHM construction of the instanton moduli space~\cite{Nek}. The $k$ instanton partition 
function can represented by  contour integrals w.r.t. certain ADHM eigenvalues $\phi_I$
$$ 
Z_{k}=\sum_{|\vec{\lambda}|=k} Z_{\vec{\lambda}} = \oint \prod_{I=1}^k 
\Big[\frac{\epsilon_1+\epsilon_2}{2\pi\epsilon_1\epsilon_2}\frac{d\phi_I}{P(\phi_I)P(\phi_I+\epsilon_1+\epsilon_2)}\Big] \prod_{I\neq J} 
\frac{\phi_{IJ}(\phi_{IJ}+\epsilon_1+\epsilon_2)}{(\phi_{IJ}+\epsilon_1)(\phi_{IJ}+\epsilon_2)}.
$$
Here we wrote a more general expression with $\epsilon_1,\epsilon_2$ equivariant parameters, which lead to the 
Seiberg-Witten solution under the identification $\hbar=\epsilon_1=-\epsilon_2$ (which we also assume in this 
paper). Moreover $\phi_{IJ}=\phi_I-\phi_J$ and
\be
P(x) = \prod_{l=1}^n (x-a_l).  \label{poly-P}
\ee 
One can now introduce the density of eigenvalues
$$
\rho(x) = \epsilon_1\epsilon_2 \sum_{I=1}^k \delta(x-\phi_I),
$$
in terms of which the measure is approximated for small $\epsilon_1,\epsilon_2$ by the saddle point method 
as
$$
\Lambda^{2kn} Z_k \sim \exp \Big(\frac{1}{\epsilon_1\epsilon_2}{\bf E}_{\Lambda}[\rho] \Big)
$$
with $\rho$ the saddle point of the functional
\be
{\bf E}[\rho] = -\int^{p.v.} dx\,dy\frac{\rho(x) \rho(y)}{(x-y)^2} -2\int dx\,\rho(x) \log P(x).  \label{Erho}
\ee

Moreover, under the identification 
$$
\rho(x) = f(x) - \sum_{l=1}^n |x-a_l|,
$$
the above functional is equivalent to the functional 
\bea
\mathcal{E}[f] & = & \frac{1}{4} \int^{p.v.}_{y<x} dx\,dy\, f''(x) f''(y)(x-y)^2 \Big(\log(x-y) - 
\frac{3}{2}\Big) = \nonumber \\
& = & -\frac{1}{2} \int_{x<y}^{p.v.} dx\, dy\, (N + f'(x))(N-f'(y)) \log(y-x),  \label{var4d} 
\eea
whose stationary point corresponds to the partition which arises as a limit shape $f_*$ of Young diagrams in 
(\ref{Nekrasov-PF}). This limit shape partition for $SU(n)$ theory is given by the function $f(x)$ of the form
$$
f(x) = \sum_{l=1}^n f_l(x-a_l).
$$
In fact, the stationary point $f_*$ of the functional $\mathcal{E}[f]$ is related to the function $\varphi(x)$ 
\be
f'_*(x) = \textrm{Re}\, \varphi(x),   \label{f-phi} 
\ee
and this is $\varphi(x)$ which explicitly encodes the Seiberg-Witten curve. More precisely, 
$\varphi(x)=\Phi(x+i0)$, where $\Phi$ is a certain conformal mapping described in detail in \cite{Nek-Ok}, such 
that the so-called bands of $\Phi$ determine the cuts of the Seiberg-Witten curve. 
The above construction was extended in \cite{nek-shad}, where 
the limit shape equations where cast in the form of the eigenvalue distribution
for all classical gauge groups.

%*******************************************************
%*******************************************************

\subsection{5-dimensional generalization}

So far we considered the 4-dimensional theory. It can be regarded as a limit of a 5-dimensional gauge 
theory compactified on a circle of circumference $\beta$. For finite $\beta$ the quantities in 5-dimensional 
theory, such as the partition function, correlation functions, or the spectral curve, are trigonometric analogues of 
the corresponding 4-dimensional quantities. In particular the partition function reads  \cite{Nek,Nek-Ok,NY-I,NY-II}
\bea
Z^{SU(n)}_{5d} & = & \sum_{\vec{\lambda}=(\lambda^{(1)},\ldots,\lambda^{(n)})} (\beta\Lambda)^{2n 
|\vec{\lambda}|} Z_{\beta,\vec{\lambda}}, \label{Nekrasov-PF-5d} \\
Z_{\beta,\vec{\lambda}} & = &
\prod_{(l,i)\neq (k,j)} \frac{\sinh \frac{\beta\hbar}{2}(p_{lk}+\lambda^{(l)}_i - \lambda^{(k)}_j + j - 
i)}{\sinh \frac{\beta\hbar}{2}(p_{lk}+j-i)},
\eea
where $|\vec{\lambda}|=\sum_{l=1}^n |\lambda^{(l)}|$. Here and in what follows we often use the (quantized) 
values $p_{l} = a_{l}/\hbar$, and $p_{lk} = p_l - p_k$.

This theory is again characterized by a Riemann surface and a meromorphic differential. 
The corresponding limit shape can be found from the variational problem for the following functional \cite{Nek-Ok}
\be
\mathcal{E}_{\beta}[f] = \int_{x<y}^{p.v.} dx\, dy\, (N + f'(x))(N-f'(y)) \log\big(\frac{2}{\beta} \sinh 
\frac{\beta|y-x|}{2}\big),   \label{var5d}
\ee
which is a direct generalization of (\ref{var4d}). In section \ref{ssec-curve5d} we will show 
that the corresponding curve can be found in a complementary way in the matrix model formalism, 
using the Migdal-Muskhelishvili formula for the resolvent. The relation of 5-dimensional gauge theory to topological string on
non-compact Calabi-Yau spaces will also be discussed in section \ref{sec-top-string}.

%*******************************************************
%*******************************************************

\section{Matrix model for 4-dimensional Seiberg-Witten theory}   \label{sec-4d}

In this section we derive matrix models for 4-dimensional Seiberg-Witten
theory. Our strategy is to introduce an auxiliary dependence on a size $N$ of
two-dimensional partitions in the Nekrasov partition function, and replace the sum
over these partitions by a sum over $N\times N$ matrices using the techniques
presented in \cite{eynard-planch}. This leads to a matrix model with a very complicated potential,
which includes such terms as logarithms of the $\Gamma$-function. However the
result we are after is rederived in the $N\to\infty$ limit, which turns into
the ordinary large $N$ limit in the matrix model formalism. In this limit we
get a matrix model with a smooth, (relatively) nice and (relatively) familiar potential. We also discuss how the
Seiberg-Witten curve arises as a spectral curve of this matrix model.

\subsection{The Nekrasov partition function as a matrix model}   \label{sec-rewrite-4d}

Our aim is to rewrite the partition function of the Seiberg-Witten theory (\ref{Nekrasov-PF}) as a matrix model. 
For simplicity, we consider first $SU(2)$ case with $n=2$. We denote $\lambda=\lambda^{(1)}$, 
$\mu=\lambda^{(2)}$, and introduce
\be
h_i = \lambda_i - i + N + p_1,  \qquad \qquad k_i = \mu_i - i + N + p_2,    \label{hk}
\ee
where we can assume $p_1<<p_2$, so that
\be 
k_1>k_2>\ldots >k_N \geq p_2 > h_1>h_2\ldots > h_N \geq p_1.   \label{order-h-k}
\ee
In terms of the variables $h_i,k_j$ in the above range the $SU(2)$ partition function can be rewritten as
$$
Z^{SU(2)} \sim \sum_{(h_i),(k_j)} \Lambda^{2n\sum_i (h_i+k_i)} \Big(\prod_{i<j}\frac{h_i - h_j}{j-i} \Big)^2 \Big(\prod_{i<j}\frac{k_i - 
k_j}{j-i} \Big)^2 \Big(\prod_{i,j}\frac{h_i - k_j}{p_{12} + j-i}  \Big)^2\ .
$$
The products in the above expression can be reorganized so that only
non-trivial $h_i, k_i$ with  $i=1,\ldots,N$ occur explicitly. Furthermore the
form of (\ref{hk}) leads for $i>N$ to many cancellations. Therefore we get for the first product above  
\bea
\prod_{i<j}\frac{h_i - h_j}{j-i} & = & \Big(\prod_{1\leq i<j \leq N} (h_i - h_j) \Big) \Big(\prod_{i=1}^N 
\prod_{j=1}^{\infty}  \frac{h_i + j - p_1}{j}\Big) = \nonumber \\
& = & \Big(\prod_{1\leq i<j \leq N} (h_i - h_j) \Big) \Big(\prod_{i=1}^N  \frac{1}{(h_i-p_1)!}\Big)   \nonumber 
%\\
%\prod_{i<j}\frac{k_i - k_j}{j-i} & = & \Big(\prod_{1\leq i<j \leq N} (k_i - k_j) \Big) \Big(\prod_{i=1}^N  
%\frac{1}{(k_i-p_2)!}\Big)   \nonumber \\
\eea
and the analogous expression for the second product. The third product
involves differences $(h_i-k_j)$ where  the index $i$ can be equal, smaller or greater than $j$. 
Similar cancellations as above lead  to
\bea
\prod_{i=j=1}^{\infty} \frac{h_i - k_j}{p_{12}} & = & \Big(\prod_{i=1}^N \frac{h_i - k_i}{p_{12}} \Big)
 \Big(\prod_{i=N+1}^{\infty} \frac{p_{12}}{p_{12}}\Big) = \frac{1}{p_{12}^N} \prod_{i=1}^N (h_i - k_i)  
\nonumber \\
\prod_{i<j}\frac{h_i - k_j}{p_{12} + j - i} & = & (p_{12}!)^N \Big(\prod_{1\leq i<j \leq N} (h_i - k_j) \Big) 
\Big(\prod_{i=1}^N  \frac{1}{(h_i-p_2)!}\Big)   \nonumber \\
\prod_{i>j}\frac{h_i - k_j}{p_{12} + j - i} & = & (-p_{12}+1)_{p_{12}} \Big(\prod_{1\leq j<i \leq N} (h_i - k_j) 
\Big) \Big(\prod_{j=1}^N  \frac{1}{(k_j-p_1)!}\Big)   \nonumber \\
\eea
respectively. 

Collecting all above contributions we observe that each difference $(h_i-h_j)$, $(k_i-k_j)$ and $(h_i-k_j)$ 
appears twice. There are also terms with factorials involving differences with $p_1$ and $p_2$, which appear for 
both $h_i$ and $k_j$ on the same footing. We can therefore introduce one set of variables
\be
l_{i=1,\ldots,2N} = (k_1,\ldots,k_N,h_1,\ldots,h_N).
\ee
In terms of $l_i$ the partition function can be written in a compact form as
\bea
Z^{SU(2)} & \sim &  N!^2 \sum_{l_1>l_2>\ldots>l_N \geq p_2 > l_{N+1} > \ldots > l_{2N}\geq p_1} 
\prod_{1\leq i < j \leq 2N} (l_i - l_j)^2 \prod_{i=1}^{2N} 
\frac{\Lambda^{2n l_i}}{(l_i-p_1)!^2 (l_i-p_2)!^2} = \nonumber \\
& = & \sum_{l_1,\ldots,l_N \geq p_2 >l_{N+1}, \ldots, l_{2N}\geq p_1} \prod_{1\leq i < j \leq 2N} (l_i - l_j)^2 
\prod_{i=1}^{2N} \frac{\Lambda^{2n l_i}}{(l_i-p_1)!^2 (l_i-p_2)!^2},   \label{Z-prod}
\eea
which is normalized with respect to the constant factors like $p_{12}^2$ and $N!^2$; they are irrelevant for 
the further discussion and drop out anyway in correlation functions, thus can be safely skipped. 
In case when negatives quantities arise under the factorials in the above product, 
we replace these factorials by $\Gamma$ functions, which 
provides the relevant analytic continuation.
%The above manipulations appear correct only for negative $p_l$, 
%due to arising factorials. However we rewrite these results 
%below in terms of $\Gamma$ functions, which provides 
%analytic continuation for arbitrary values of $p_l$.  

The expression (\ref{Z-prod}) already has a flavour of the matrix model, with the first product being 
the Vandermonde determinant. To write it explicitly as a matrix integral, we repeat the trick from 
\cite{eynard-planch} and introduce the function
$$
f(x) = -x \G(-x) \G(x) e^{-i \pi x} = \frac{\pi e^{-i\pi x}}{\sin(\pi x)},
$$
which has simple poles at all integer values of the argument. 
Upon integration along the contour $C$ encircling $[ p_1,\infty [$ 
part of the real axis, this function can be used to pick up all 
integer values of $h_i\in [ p_1,\infty [$. Similarly, one would have to 
introduce another contour $C'$ encircling $[ p_2,\infty [$ 
half-line to pick up relevant values of $k_i$. 
In fact, we can use the same contour $C$ for both $h_i$ and $k_j$,
because additional $k_j\in ]p_1,p_2 [$ picked up by the latter integral contribute zero
(for such $k_j$, $\Gamma(k_j-p_2+1)^{-2}$ has a double zero which cancels the simple pole of $f(k_j)$).
This leads to the following expression for (\ref{Z-prod})
$$
Z^{SU(2)} =  \oint_{C^{2N}} dx_1\cdots dx_{2N} \prod_{1\leq i<j \leq 2N} (x_i-x_j)^2 \prod_{i=1}^{2N} 
\frac{f(x_i)\,\Lambda^{2n x_i}}{\G(x_i-p_1+1)^2 \G(x_i-p_2+1)^2} =
$$
\be
=  \int_{Mat_{2N}(C)} \mathcal{D}M e^{-\frac{1}{\hbar} \Tr V(M)},   \label{SWmatrix}
\ee
where $M\in Mat_{2N}(C)$ is $2N\times 2N$ matrix with eigenvalues in a set
$C$. Choosing real values of $p_l$ we can in fact reduce the contour to the
real axis in the large $N$ limit. It is convenient to rescale $x\mapsto 
x/\hbar$, and in terms of this rescaled variable the potential reads
\bea
\frac{1}{\hbar} V(x) & = & -(2n \log\Lambda) x -\log \, f\big(\frac{x}{\hbar}\big) + \sum_{l=1,2} \log \G(x 
\hbar^{-1}-p_l+1)^2 = \nonumber \\
& = & -(2n\log\Lambda)x -\log\big(\frac{x}{\hbar}\big) -\log \,\G\big(-\frac{x}{\hbar}\big) -\log\,\G\big(\frac{x}{\hbar}\big) + 
\frac{i\pi x}{\hbar} + \nonumber \\
& & + 2\sum_{l}\Big(\log\big(x\hbar^{-1}-p_l\big) + \log\,\G\big(x\hbar^{-1}-p_l\big) \Big).   \label{SW-V} 
\eea
This expression looks quite complicated. Nonetheless, we will see that in the large $N$ limit it simplifies enormousely.

Now it also becomes clear that in generalization to arbitrary $SU(n)$ theory we need to introduce $l_i$ for 
$i=1,\ldots,nN$, which is a concatenation of strings of $N$ eigenvalues corresponding to matrices corresponding 
to each partition $\lambda^{(l)}$. In the last sum of (\ref{SW-V}) we also need to take the range of $l$ from 
$1$ to $n$.

In what follows also the derivative of this potential plays an important role. It involves the logarithmic 
derivative of the $\Gamma$ function, i.e. the digamma function
\be
\psi(x) = \frac{d}{dx} \log\,\G(x) = \frac{\G'(x)}{\G(x)}.
\ee 
This function shares a number of interesting properties, for example
$$
\psi(x+1) = -\gamma - \sum_{n=1}^{\infty}\big(\frac{1}{x+n}-\frac{1}{n}\big) = \log\,x + \frac{1}{2x} - 
\sum_{i=1}^{\infty} \frac{B_{2i}}{2i x^{2i}},
$$
where $\gamma$ is Euler-Mascheroni constant and $B_{2i}$ are Bernoulli numbers. From this we also deduce
$$
\psi(x) = -\frac{1}{2x} + \log\, x - \sum_{i=1}^{\infty} \frac{B_{2i}}{2i x^{2i}}.
$$
From these expressions we find
\bea
V'(x) & = & -2n\hbar\log\Lambda + \sum_l \Big(\frac{1}{x\hbar^{-1}-p_l} + 2\log\big(\frac{x}{\hbar}-p_l\big) - \sum_{i=1}^{\infty} \frac{B_{2i}}{i(x\hbar^{-1}-p_l)^{2i}} \Big) = \nonumber \\
& = & -2n(\log \hbar +\hbar\log\Lambda) + \sum_l \Big(\frac{\hbar}{x-a_l} + 2\log(x-a_l) - \sum_{i=1}^{\infty} \frac{\hbar^{2i} B_{2i}}{i(x-a_l)^{2i}} \Big).       \label{SW-Vprim}
\eea

It is important to stress that the potential (\ref{SW-V}), and so of course its derivative above, explicitly do 
not depend on the size of matrices $N$. \footnote{Some complication due to explicit $N$ dependence in eq. (2.23) in \cite{eynard-planch} arouse due to higher $t_k$ couplings, which we do not consider here.}
%This is the necessary condition for the consistency of the solution 
%\cite{eynard-planch,CGMPS}. In particular in \cite{eynard-planch}, 
%in $n=1$ case, a dependence on $N$ did 
%arise from higher Casimir terms. We do not consider 
%such terms in the present case, so the consistency of the result is automatic. 
For finite $N$, similarly as discussed in \cite{eynard-planch}, 
there could be at most exponentially small corrections in $N$, which vanish in the large $N$ limit 
we consider in what follows.

\begin{figure}[htb]
\begin{center}
\includegraphics[width=0.4\textwidth]{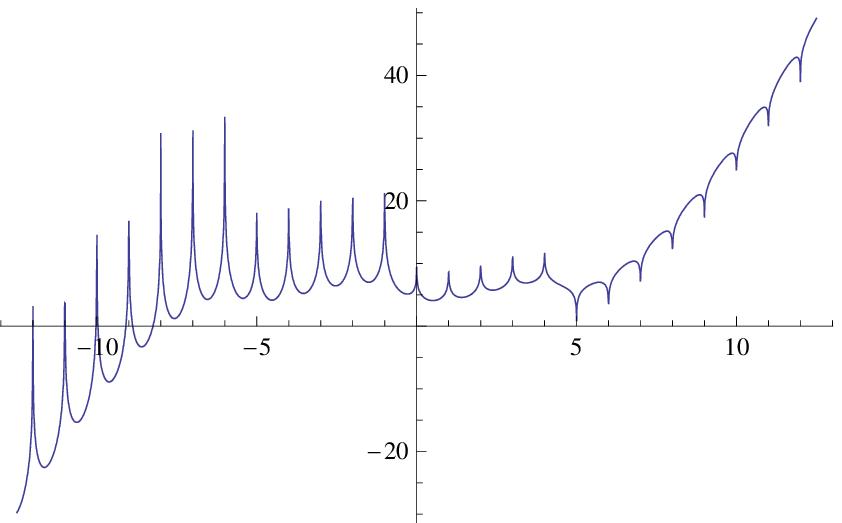} $\qquad$
\includegraphics[width=0.4\textwidth]{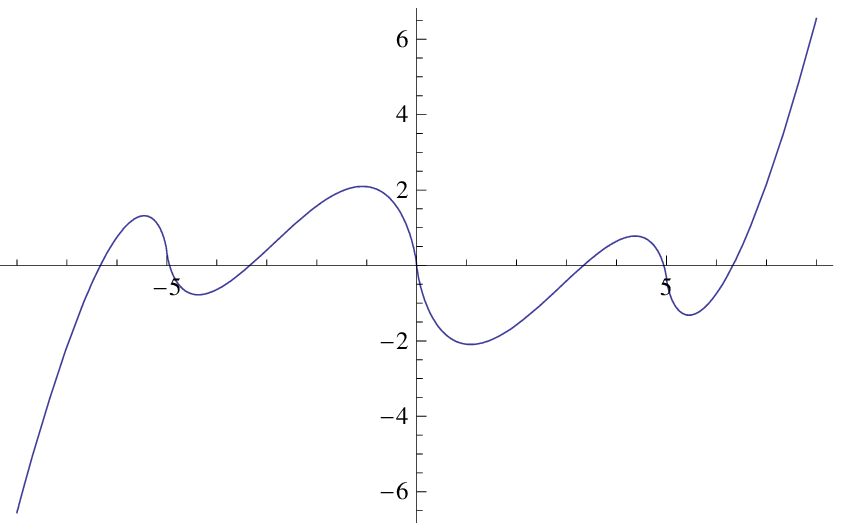}
\begin{quote}
\caption{Matrix model potential for $SU(3)$ Seiberg-Witten theory.
Left: the potential corresponding to the exact expression (\ref{SW-V}) for
finite $N$; complicated spikes arise from $\log\Gamma$ terms. 
Right: the potential in the large $N$ limit, given by (\ref{SW-V4d}).
Three minima give rise to three cuts and genus 2 Seiberg-Witten curve.} \label{fig-V}
\end{quote}
\end{center}
\end{figure}

The above expression can also be integrated and yields
\be
V(x) = tx + 2\sum_l \Big( (x-a_l)\log(x-a_l) - (x-a_l) \Big) + \mathcal{O}(\hbar),   
\ee
where coefficients of linear terms are encoded in the constant $t$. We stress that so far we considered a finite value of $N$ corresponding to the number of rows in partitions which appear in the Nekrasov partition function (\ref{Nekrasov-PF}). Therefore the original Nekrasov partition function is automatically obtained in the large $N$ limit for the matrix model expression (\ref{SWmatrix})
\be
N\to\infty,\qquad \qquad \hbar\to 0,\qquad \qquad \hbar N = const.    \label{largeN}
\ee
In this limit all the terms of order $\mathcal{O}(\hbar)$ vanish, so we conclude that the $SU(n)$ Seiberg-Witten theory can be formulated in terms of the 1-matrix model with the potential
\be
V^{4d}(x) = tx + 2\sum_{l=1}^n \Big( (x-a_l)\log(x-a_l) - (x-a_l) \Big),   \label{SW-V4d}
\ee
where $t$ in the linear term encodes in particular the scale $\Lambda$. This potential is shown in figure \ref{fig-V}. For $SU(n)$ theory it clearly has $n$ minima to which we can associate $n$ cuts, giving rise to the Seiberg-Witten spectral curve of genus $(n-1)$. 

%*******************************************************
%*******************************************************

\subsection{Seiberg-Witten curve as a spectral curve}   \label{ssec-curve}

Let us recall some basic facts concerning matrix models. For the matrix integral
$$
Z = \int \mathcal{D}M e^{-\frac{1}{\hbar}\Tr V(M)}
$$
over $N\times N$ matrices, the eigenvalues are subject to the effective potential
$$
S_{eff}(\lambda) = 2\sum_{i<j} \log |\lambda_i-\lambda_j| +
\frac{1}{\hbar}\sum_{i=1}^N V(\lambda_i). 
$$
Replacing discrete eigenvalues by a continuous eigenvalue density $r(x)$
$$
\frac{1}{N}\sum_{i=1}^N h(\lambda_i) \mapsto \int h(x)r(x)\,dx
$$
leads to the effective functional
\be
S_{eff}[r] = \int_{x\neq y} dx\,dy\,r(x)r(y) \log |x-y| -\int dx\, r(x) V(x) ,    \label{Seff}
\ee
and the spectral curve of the matrix model is encoded in the discontinuities of the resolvent 
which leads to the minimization of this functional.

Let us now introduce $R(x)$ such that
\be
R'(x) = r(x),  \label{Rprim-r}
\ee
and integrate (\ref{Seff}) by parts to get
\be
S_{eff}[R] = \int_{x\neq y} dx\,dy\,\frac{R(x)R(y)}{(x-y)^2} + \int dx\, R(x) V'(x).    \label{SeffR}
\ee
We focus now on the matrix model for Seiberg-Witten theory (\ref{SWmatrix}), with a potential whose derivative 
is given in (\ref{SW-Vprim}). Let us stress that the matrix 
model (\ref{SWmatrix}) still depends on both $N$ (size of a matrix) and $\hbar$. 
Upon extracting a spectral curve corresponding to the Seiberg-Witten theory we should again take the limit (\ref{largeN}) in which all terms in (\ref{SW-Vprim}) proportional to $\hbar$ are neglected. The only terms which survive are
\be
V^{4d}{'}(x) = 2 \sum_l \log(x-a_l) = 2\log P(x),   \label{skipB}
\ee
where $P(x)=\prod_{l=1}^n (x-a_l)$ coincides with (\ref{poly-P}). Plugging this into (\ref{Seff}) we get 
\be
S_{eff}[R] = \int_{x\neq y} dx\,dy\,\frac{R(x)R(y)}{(x-y)^2} +  2 \int dx\, R(x) \log P(x).   \label{SeffRbis}
\ee
We observe that this functional is identical to (\ref{Erho}) if we identify $R(x)=\rho(x)$. From the analysis in 
\cite{Nek-Ok} we know that the stationary point of (\ref{Erho}) leads to the Seiberg-Witten curve encoded in the 
derivative of the limiting profile (\ref{f-phi}). In our case we also need to take the derivative (\ref{Rprim-r}) 
of $R(x)$ to get $r(x)$ corresponding to the eigenvalue density. We conclude that the stationary point of 
(\ref{Seff}) -- encoding the spectral curve of our matrix model -- corresponds to the same Seiberg-Witten curve 
as the one derived in \cite{Nek-Ok}. 

The above argument relies on the form of the functional (\ref{Erho}) considered by
Nekrasov-Okounkov. There is however a way to obtain the spectral curve
directly within the matrix model formalism. In more generality this works also
for the 5-dimensional matrix model which we introduce in the next section. One
can therefore derive the spectral curve for the 5-dimensional theory according to the procedure presented
in the subsection \ref{ssec-curve5d}, and then recover the curve discussed
above in the 4-dimensional limit.

%*******************************************************
%*******************************************************

%*******************************************************

\section{Matrix model for 5-dimensional gauge theory}    \label{sec-5d}

In this section we generalize the construction of matrix models for 4-dimensional gauge theories in 
section~\ref{sec-4d}  to a canonical class of 5-dimensional gauge theories. 
This construction will be further slightly generalized in section~\ref{sec-top-string}.

\subsection{5-dimensional partition function as a matrix model}

We now wish to introduce a matrix model for the 5-dimensional $SU(n)$ theory with a partition function given in 
(\ref{Nekrasov-PF-5d}). Similarly as before, to start with we consider $SU(2)$ case with $n=2$. We denote 
$\lambda=\lambda^{(1)}$ and $\mu=\lambda^{(2)}$, which are partitions with no more than $N$ rows, and similarly as in 
(\ref{hk}) introduce
\be
h_i = \lambda_i - i + N + p_1,  \qquad \qquad k_i = \mu_i - i + N + p_2.    \label{hk-5d}
\ee
Moreover we assume $p_1<<p_2$, so that
\be 
k_1>k_2>\ldots >k_N \geq p_2 > h_1>h_2\ldots > h_N \geq p_1.   \label{order-h-k-5d}
\ee

We also introduce $q=e^{-g_s}=e^{-\beta \hbar}$ and the standard $q$-deformed notation
$$
[ h ] = q^{-h/2} -q^{h/2},  \qquad \qquad [ h ]!  =  \prod_{i=1}^h (q^{-i/2} - q^{i/2}). 
$$

The $SU(2)$ partition function now takes the form
$$
Z^{SU(2)}_{5d} \sim \sum_{(h_i),(k_j)} (\beta\Lambda)^{2n\sum_i (h_i+k_i)} 
\Big(\prod_{i<j}\frac{[h_j - h_i]}{[i-j]} \Big)^2 \Big(\prod_{i<j}\frac{[k_j 
- k_i]}{[i-j]} \Big)^2 \Big(\prod_{i,j}\frac{[k_j-h_i]}{[i-j-p_{12}]}  \Big)^2,
$$
with $(h_i),(k_j)$ satisfying the condition (\ref{order-h-k-5d}). These terms can be rewritten analogously as in 
section \ref{sec-rewrite-4d}. For example
$$
\prod_{i<j} \prod_{i<j}\frac{[h_j - h_i]}{[i-j]} = \Big(\prod_{1\leq i<j \leq N} [h_j - h_i] \Big) 
\Big(\prod_{i=1}^N \frac{1}{[h_i-p_1]!}\Big),  
$$
and similarly for other terms. Therefore, introducing one set of variables
\be
l_{i=1,\ldots,2N} = (k_1,\ldots,k_N,h_1,\ldots,h_N),
\ee
we get
\be
Z^{SU(2)}_{5d} = \sum_{(l_i)_{i=1,\ldots,2N}} \Big(\prod_{1\leq i<j \leq 2N} [l_i - l_j]^2 \Big) 
\Big(\prod_{i=1}^{2N} \frac{(\beta\Lambda)^{2n l_i}}{([l_i-p_1]! \, [l_i-p_2]!)^2}\Big).
\ee

To write this expression as a matrix model integral we again take advantage of a few facts discussed in 
\cite{eynard-planch}. First of all, we introduce the following notation for the quantum dilogarithm
$$
g(x) = \prod_{i=1}^{\infty} (1-x^{-1}q^i).
$$
It vanishes $g(q^h)=0$ for $h$ a positive integer, and at such points its derivative is
$$ 
g'(q^h) = - \frac{g(1)^2 e^{i\pi h} q^{-h(h-1)/2}}{q^h(1-q^h)g(q^{-h})}.
$$
Therefore the following function has simple poles with residue 1 for $x=q^h$ with $h\in\mathbb{N}$
$$
f(x) = -\frac{g(1)^2 e^{-\frac{i\pi}{g_s}\log x} e^{\frac{(\log x)^2}{2g_s}}}{(1-x)\sqrt{x}g(x)g(x^{-1})}.
$$
Let us also note that for $x=q^l$
\bea
[l]! & = & q^{-l(l+1)/4} \frac{g(1)}{g(x^{-1})}, \nonumber \\
\frac{1}{([l-p]!)^2} & = & \frac{g(x^{-1}q^p)^2}{g(1)^2} \exp\big(\frac{1}{2\log q} \log(xq^{-p}) 
\log(xq^{-p+1})\big). \nonumber 
\eea

In what follows we use the notation 
\be
x_i = q^{l_i} = e^{-g_s l_i} = e^{u_i},  \label{xu}
\ee
so that $x_i$ is a cylindrical coordinate. Now we can write
\bea
Z^{SU(2)}_{5d} & \sim & \sum_{(l_i)_{i=1,\ldots,2N}} \prod_{1\leq i<j \leq 2N} \big(q^{l_i} - q^{l_j}\big)^2  
\prod_{i=1}^{2N} \frac{(\beta\Lambda)^{2n l_i}  q^{(1-2N)l_i}}{([l_i-p_1]!)^2 ([l_i-p_2]!)^2} = \nonumber \\
& = & \int dx_1\cdots dx_{2N} \prod_{1\leq i < j \leq 2N} (x_i-x_j)^2 \times  \\
& & \qquad \qquad \times \prod_{i=1}^{2N} \frac{(\beta\Lambda)^{-\frac{2n}{g_s} \log x_i} f(x_i)}{x_i^{2N-1}}\prod_{l=1}^{n=2} 
\Big[\frac{g(x_i^{-1}q^{p_l})^2}{g(1)^2}e^{\frac{\log(x_i q^{-p_l}) \log(x_i q^{-p_l+1})}{2\log q}}\Big]. \nonumber
\label{5deigenvalue}\eea
The deformed Vandermonde determinant can also be written as
\be
e^{(1-N) \sum_i \log x_i} \prod_{i<j} (x_i-x_j)^2 = \prod_{i<j} \Big(2 \sinh \frac{\log x_i - \log x_j}{2}\Big)^2 \equiv \Delta_q^2.   \label{Van-sinh}
\ee
With the deformed measure 
$$
\mathcal{D}M = \Delta_q^2  \prod_i d u_i = \Delta_q^2 e^{-\sum_i \log x_i} \prod_i dx_i
$$
we can rewrite the above partition function as a matrix model expression, 
which in a straightforward way generalizes to arbitrary $SU(n)$ gauge group
$$
Z^{SU(n)}_{5d} = \int_{M_{nN}(C_q)} \mathcal{D}M e^{-\frac{1}{g_s} \Tr V_{5d}(M)},
$$
with the deformed integration contour $C_q$ is a circle of radius $r\in]1,q^{-1}[$. 
The matrix model potential takes the form
\bea
\frac{1}{g_s} V_{5d}(x) & = & \Big(\frac{2n \log (\beta\Lambda)}{g_s} +  (n-1)N - 1\Big)\log x -2\log g(1) + \log g(x)+ \log g(x^{-1}) + \nonumber \\
& & + \frac{1}{2}\log x + \log (1-x) + \frac{i\pi}{g_s} \log x - \frac{1}{2g_s}(\log x)^2 -i \pi + \nonumber \\
& & + \sum_{l=1}^n\Big[2\log g(1)  -2\log g(x^{-1}q^{p_l}) - \frac{\big(\log (xq^{-p_l}) \big)^2}{2\log q} - 
\frac{1}{2} \log(xq^{-p_l})\Big].  \label{V-5d} 
\eea

Similarly as in 4-dimensional case, this expression also simplifies in the large $N$ limit. 
To turn the potential into a nicer form, we again differentiate it and then integrate.
First we note \cite{eynard-planch} that
$$
\log g(x) = -\frac{1}{g_s} \sum_{m=0}^{\infty} \Li_{2-m}(x^{-1}) \frac{B_{m}g_s^m}{m!},
$$
where $B_m$ are Bernoulli numbers. The polylogarithm is defined as 
$$
\Li_m(x) = \sum_{i=1}^{\infty} \frac{x^i}{i^m},
$$
and it has the following properties
\bea
\Li_n'(x) & = & \frac{\Li_{n-1}(x)}{x}, \nonumber \\
\Li_{m}(x) & = & (-1)^{m+1} \Li_m(x^{-1}), \qquad \quad \textrm{for}\ m<0 \nonumber \\
\Li_0(x) & = & \frac{x}{1-x},\nonumber \\
\Li_1(x) & = & -\log(1-x).
\eea
The following useful relations follow
\bea
\frac{g'(x)}{g(x)} & = & \frac{1}{g_s x} \sum_{m=0}^{\infty} \Li_{1-m} (x^{-1}) \frac{B_m g_s^m}{m!}, \nonumber 
\\
x \frac{d}{dx}\big(\log g(x) + \log g(x^{-1})\big) & = & \frac{\log x}{g_s} - \frac{1}{2}\frac{x+1}{x-1} - 
\frac{i\pi}{g_s}. \nonumber
\eea

\begin{figure}[htb]
\begin{center}
\includegraphics[width=0.4\textwidth]{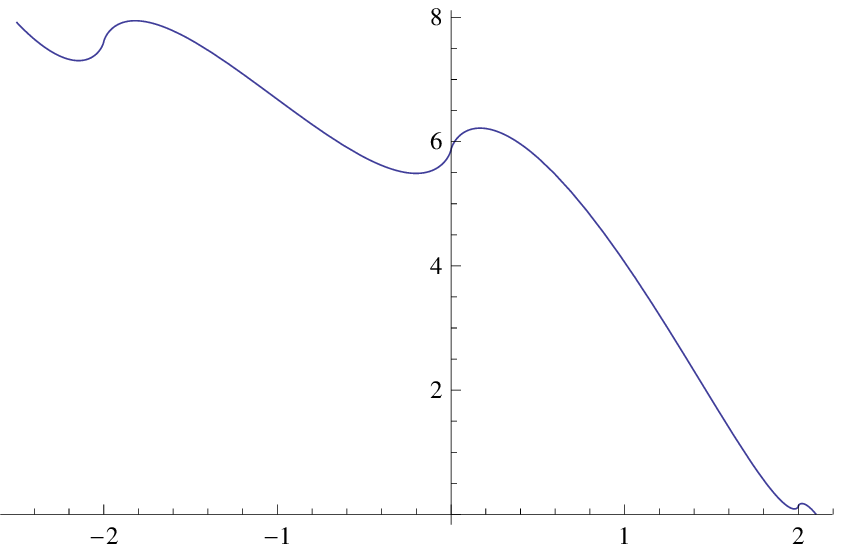} $\qquad$
\includegraphics[width=0.4\textwidth]{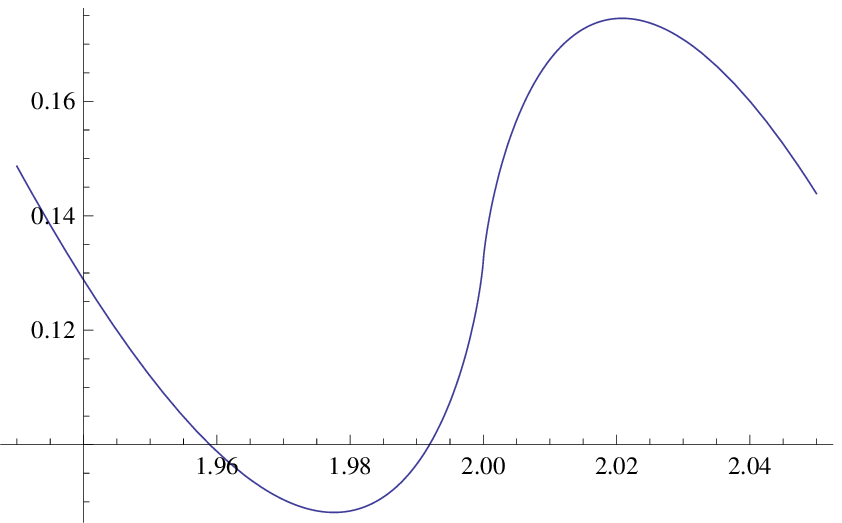}
\begin{quote}
\caption{Matrix model potential for 5-dimensional $SU(3)$ Seiberg-Witten theory.
Left:  the potential in the large $N$ limit, given by (\ref{SW-V5d}). Three
minima give rise to three cuts and the genus 2 Seiberg-Witten curve.
Right: zooming around the rightmost minimum.
} \label{fig-V5d}
\end{quote}
\end{center}
\end{figure}

Using the above facts, after some algebra one therefore finds the derivative of the potential
$$
V_{5d}'(x) =  \frac{2n\log(\beta\Lambda) + g_s ((n-1)N-\frac{1}{2})}{x}  + 
$$
\be
+  \sum_{l=1}^n \Big[\frac{g_s}{x-q^{p_l}}  -\frac{2}{x} \log\big((xq^{-p_l})^{-\hf} - (xq^{-p_l})^{\hf}\big) + \frac{2}{x}\sum_{m=1}^{\infty} \Li_{1-2m}(xq^{-p_l}) \frac{B_{2m}g_s^{2m}}{(2m)!}  \Big].   \label{Vprim-5d}
\ee
Let us note that $\log\big((xq^{-p_l})^{-\hf} - (xq^{-p_l})^{\hf}\big)$ is the q-deformation of the 4-dimensional 
expression $\log(x-a_l)$ in (\ref{SW-Vprim}), while other terms (apart from the linear in $1/x$) are proportional to higher powers of $g_s$. The expression which we obtained can now be integrated. In particular
$\int \frac{dx}{x} \log(x^{-1/2} - x^{1/2}) = 
%\int \Big(-\frac{1}{2}\frac{\log x}{x} - \frac{\Li_1(x)}{x}  \Big)dx 
-\frac{1}{4}(\log x)^2 - \Li_2(x),$ 
so that
\be
V_{5d} = t\log x + \frac{n}{2}(\log x)^2 + 2\sum_l \Li_2(x e^{a_l}) + \mathcal{O}(g_s),
\ee
where the coefficient $t$ encodes in particular the scale $\beta\Lambda$ and terms linear in $a_l$. In the large $N$ limit
$$
N\to\infty,\qquad \qquad g_s\to 0,\qquad \qquad g_s N = const,    
$$
we can again neglect all terms $\mathcal{O}(g_s)$. We also note that introduction of the single parameter $t$ is justified, as it involves both $\log(\beta\lambda)$ and $g_s N = const$ related to each other in large $N$ dualities precisely in this way. Substituting now $x=e^u$, finally we get the matrix model
\be
Z^{SU(n)}_{5d} = \int  \mathcal{D}M e^{-\frac{1}{g_s} \Tr V^{5d}(M)} =  \int \prod_i d u_i  \prod_{i<j} \Big(2 \sinh \frac{u_i - u_j}{2}\Big)^2   e^{-\frac{1}{g_s} \sum_i V^{5d}(u_i)}
\ee
with the potential
\be
V^{5d}(u) = t u + \frac{n}{2} u^2 + 2\sum_{l=1}^n \Li_2(e^{u+a_l}).   \label{SW-V5d}
\ee
The coefficient $n$ in the quadratic term can be interpreted as redefining the coupling in front of the potential as
$$
g_s \mapsto \hat{g_s} = \frac{g_s}{n}
$$
(and rescaling other terms in the potential by $n$). It is consistent with the same rescaling found in a related matrix model in \cite{lens-matrix}, which we discuss in section \ref{sec-models}.
The potential $V^{5d}$ is shown in figure \ref{fig-V5d}.

\subsection{5-dimensional spectral curve}   \label{ssec-curve5d}

Finding the spectral curve for 5-dimensional theory also requires the 't Hooft
limit, in which all terms in the potential (\ref{Vprim-5d}) proportional to
positive powers of $g_s$ vanish, so that 
$$
xV'_{5d}(x) \ \longrightarrow  \ -2\sum_{l=1}^n \log\big((xq^{-p_l})^{-\hf} - (xq^{-p_l})^{\hf}\big) = -2 \log 
P_q(y), 
$$
$$
 \textrm{for} \quad P_q(y) = \prod_{l=1}^n [y-p_l],
$$
with $x=q^y$. This is indeed the trigonometric counterpart of the 4-dimensional case (\ref{skipB}). Accordingly the 
5-dimensional curve will be the trigonometric counterpart of the 4-dimensional Seiberg-Witten curve, in 
agreement with the relation between 4-dimensional and the 5-dimensional variational problems (\ref{var4d}) and 
(\ref{var5d}). 

Apart from arguments which rely on the form of effective functionals (which
were also employed in section \ref{ssec-curve}), we can in fact derive the
form of the 5-dimensional spectral curve using purely matrix model technology,
i.e. determining the resolvent and the eigenvalue density in the large $N$
limit. This method could also have been be applied explicitly in section
\ref{ssec-curve}, however it is more convenient to consider the general
5-dimensional case and then get the 4-dimensional results in the gauge theory limit.
For a general potential in the multi-cut case the resolvent $\omega(z)$
is given by the Migdal-Muskhelishvili formula \cite{migdal}. It has the following form in
terms of the periodic variable $x$
\be
\omega(z) = \frac{1}{2\pi i} \oint_C dx \frac{V'(x)}{x(z-x)}
\sqrt{\prod_{k=1}^{2n} \frac{z-x_k}{x-x_k}},  \label{resolvent5d}
\ee
where $C=C_1\cup\ldots\cup  C_n$ is a union of $n$ cuts supported on intervals
$C_k=[x_{2k-1},x_{2k}]$. The $2n$ endpoints of the cuts $x_k$ must be suitably
chosen. A set of $(n+1)$ conditions for these endpoints stems from the limiting behaviour of the
resolvent at infinity $\omega(z)\sim 1/z$. Another $(n-1)$ conditions can be
chosen in a way which is appropriate for a given physical problem. In our case
it is natural to fix $(n-1)$ filling fractions, which are determined by the
A-periods around the cuts. 

To compute the resolvent (\ref{resolvent5d}) one can follow the solution
presented in \cite{CGMPS}, where the case corresponding to $n=1$ (which we
discuss also in section \ref{ssec-bundleP1}) was analyzed in detail. 
This solution relies on the chiral ansatz and the arctic
circle property of two-dimensional partitions which we discussed above. We
assume that the eigenvalue density $\rho(z)$ is non-trivial only in certain
intervals, independently for each set of partitions summed over in the Nekrasov
partition function. The parameters $a_l$ in the potential (\ref{SW-V5d}) can
be absorbed into the positions of the endpoints of these integrals, so that the resolvent becomes
\be
\omega(z) = \frac{1}{2\pi i} \oint_C dx \frac{n\log x + t -2\sum_{k=1}^n \log(1-xe^{y_{2k}})}{x(z-x)}
\sqrt{\prod_{k=1}^{2n} \frac{z-e^{-y_{2k-1}}}{x-e^{-y_{2k}}}}.   \label{res5d-bis}
\ee
To perform this integral one can modify the integration contour, so that it
encircles the branch cuts of all logarithms and the pole at $x=z$.

\begin{figure}[htb]
\begin{center}
\includegraphics[width=0.8\textwidth]{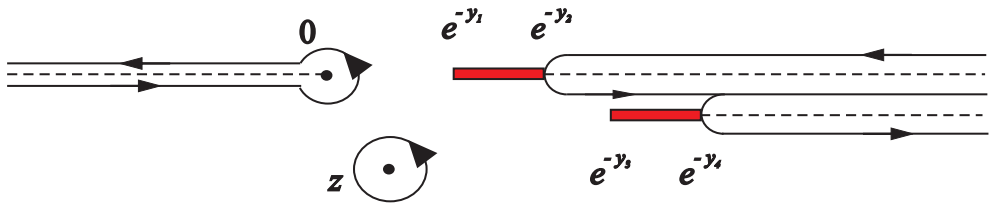} 
\begin{quote}
\caption{Integration contour used to compute the resolvent for 5-dimensional
  $SU(2)$ Seiberg-Witten theory.} \label{fig-contour}
\end{quote}
\end{center}
\end{figure}

As an example let us consider how to derive the spectral curve for the $SU(2)$
theory from our matrix model. We denote the endpoints of the two cuts as $[y_1,y_2]$ and
$[y_3,y_4]$. The integration contour in this case is shown in figure \ref{fig-contour}. 
The resolvent becomes a sum of the residue at $x=z$ and the contour integral
along the (infinitesimal) circle around $x=0$ (a part of
the contour around the branch cut of $\log x$), as well as contributions from the
integrals along parts of real axis from branch cuts of all logarithms. The
latter are given by integrals of the form
\be
\int \frac{dv}{v(z-v) \sqrt{\prod_{i=1}^4(v-e^{-y_i})}}.     \label{res-integral}
\ee
They can be expressed in terms of elliptic functions of the first and the
third kind, as explained in the appendix. In what follows we also use the
notation introduced in the appendix. After some simplifications, for $SU(2)$ theory 
the resolvent (\ref{res5d-bis}) turns out to involve three integrals of the form
(\ref{res-integral}) evaluated at the points $v=0,\infty,e^{-y_2}$ 
(the integral evaluated at $e^{-y_4}$ vanish). 
For these points the arguments of the elliptic functions $F$ and $\Pi$ are respectively 
$$
y(0)=\frac{e^{-y_4}(e^{-y_3}-e^{-y_2})}{e^{-y_3}(e^{-y_4}-e^{-y_2})}, \qquad
y(\infty)=\frac{e^{-y_3}-e^{-y_2}}{e^{-y_4}-e^{-y_2}}, \qquad
y(e^{-y_2})=1, 
$$
as well as $\delta$ and $k^2$ given in (\ref{ydeltak2}) with $a,b,c,d$ identified with $e^{-y_i}$.
Using the addition formula (\ref{addF}) twice we can replace three elliptic functions
$F(y(v),k)$ by a single one with the argument $y_{0,\infty,y_2}$ determined by using
(\ref{kat-phi}) twice. Similarly, using (\ref{addPi}) twice we get a single function
$\Pi(y_{0,\infty,y_2},\delta,k)$, but then in addition two terms $(\textrm{artanh}\,\Theta_{0,\infty})$
and $(\textrm{artanh}\,\Theta_{0,\infty,y_2})$, where $\Theta_{0,\infty}$ and $\Theta_{0,\infty,y_2}$
are determined from the formula (\ref{kat-Theta}) used twice. Altogether, the resolvent becomes
\bea
\omega(z) & = & \frac{t + 2\log z -2\log(1-ze^{y_2}) -2\log(1-ze^{y_4})}{z} +
\frac{\sqrt{\prod_i (z-e^{-y_i})}}{z} \,t e^{\frac{y_1+y_2+y_3+y_4}{2}} + \nonumber\\
& + & \frac{4\sqrt{\prod_i
    (z-e^{-y_i})}}{z \sqrt{(e^{-y_4}-e^{-y_1})(e^{-y_3}-e^{-y_2})}} \Big[ 
\Big(\frac{1}{e^{-y_3}-z} - \frac{1}{e^{-y_3}}\Big) F(y_{0,\infty,y_2}) +\nonumber  \\
& + & \frac{e^{-y_4}-e^{-y_3}}{e^{-y_3}e^{-y_4}} \Big(\Pi(y(0),\delta(0),k) + \Pi(y(\infty),\delta(0),k) - \Pi(y(e^{-y_2}),\delta(0),k)\Big) + \nonumber \\
& + &  \frac{e^{-y_4}-e^{-y_3}}{(z-e^{-y_3})(z-e^{-y_4})}\Big(\Pi(y_{0,\infty,y_2},\delta(z),k) + 
\frac{\textrm{artanh}\,\Theta_{0,\infty} + \textrm{artanh}\,\Theta_{0,\infty,y_2})}{\sqrt{\delta(z)^{-1}(\delta(z)-1)(\delta(z)-k^2)}} \Big)  \Big].\nonumber 
\eea

We discuss now the behavior of this resolvent at infinity $z\to\infty$, where it should
decay like $\sim 1/z$. Firstly, we note that the last term in the first line,
the last term $\sim e^{y_3} F(y_{0,\infty,y_2})$ in the second line, and 
the terms in the third line are of order $\sim z$. Therefore a sum of
coefficients of these terms must be zero, which gives one equation for the
endpoints of the cuts. Due to simple the same dependence on $z$, these terms in
fact cancel altogether and drop out from the expression. 
Then, the only term which is constant at infinity is the first term $\sim F(y_{0,\infty,y_2})/(e^{-y_3}-z)$ in the second line. It vanishes if
and only if the elliptic function $F(y_{0,\infty,y_2})$ vanishes, which is possible if and only
if $\sin\varphi \equiv y_{0,\infty,y_2} = 0$. One one hand, this provides another condition on the
endpoints of the cuts, due to (\ref{kat-phi}). On the other hand, then the terms in the fourth line
$\Pi(y_{0,\infty,y_2}=0,\delta(z),k)$, as well as $(\textrm{artanh}\,\Theta_{0,\infty,y_2})$ 
(whose argument is proportional to $\sin\varphi \equiv y_{0,\infty,y_2}$ via (\ref{kat-Theta})) also vanish. 
Therefore the only term which could contribute to the eigenvalue density  
(or equivalently the spectral curve) is $(\textrm{artanh}\,\Theta_{0,\infty})$ in the fourth line. 
Quite non-trivial cancellations lead to its overall prefactor being just $2/z$ 
which in particular does not depend on any $e^{-y_i}$.
Moreover, this term indeed has the square root branching as hoped for,
which arises via the dependence on $z$ through the argument of $\Theta_{0,\infty}$ 
$$
\sqrt{\delta(z)(1-\delta(z))(\delta(z) - k^2)} \sim \frac{\sqrt{\prod_i
    (z-e^{-y_i})}}{(e^{-y_4}-z)^2}\ .
$$
Indeed some algebra reveals that $\Theta_{0,\infty} = \sqrt{\prod_i
 (z-e^{-y_i})}/P(z)$, with $P(z)$ a  quadratic polynomial in $z$.

Identifying now the density with a new variable $v/z$ we get 
$$
\frac{v}{z} \equiv \rho(z) = \frac{1}{2\pi i}\big(\omega(z-i\epsilon)-\omega(z+i\epsilon)\big) \sim
\frac{2}{z} \,\textrm{artanh}\,\Theta_{0,\infty} = \frac{1}{z}\log\frac{1+\Theta_{0,\infty}}{1-\Theta_{0,\infty}}.
% \label{res-int}
$$
Exponentiating both sides we get an equation in two $\mathbb{C}^*$
coordinates, $z$ and $x\equiv e^v$. This defines a Riemann surface of genus
one  written in terms of two $\mathbb{C}^*$ coordinates $x$ and $z$, 
as expected for 5-dimensional Seiberg-Witten theory. The latter one  can be
brought to hyperelleptic form after the redefinition $\frac{x-1}{x+1}\mapsto
\tilde x$  and a rescaling of the new variable $\tilde x$ by $P(z)$, which to
the 
equation of the
curve
\be
{\tilde x}^2 = \prod_{i=1}^4 (z-e^{-y_i}).
\ee
 Three of the $y_i$'s are specified by the boundary conditions of the resolvent, and
they depend also on the modulus $t$. To get the full solution, one more
condition can be specified by imposing the A-period filling fraction. On the
other hand, this A-period can be found from the knowledge of the above curve. 

Let us stress at this point, that the appearance of integrals (\ref{res-integral}) of this
form (in particular leading to elliptic functions for $SU(2)$ theory) is a
consequence of the logarithmic branch cut
structure of the derivative of the potential $V(x)$ which appears in
(\ref{res5d-bis}). Derivatives of both the quadratic and dilogarithm terms which appear in the
potential are logarithms with branch cuts that we just need. This is therefore
non-accidental that no other (special) functions arise in the potential. 

The above derivation could of course be generalized to arbitrary $SU(n)$
theory, though technically the solution would be even more complicated. 
In this general case the well-known matrix model relations \cite{cmmv} for the filling
fractions $a_l$ and the genus zero free energy $F_0$, written in terms of the rescaled variables
$$
a_l = \frac{1}{2\pi i} \oint_{A_l} \log x \frac{dz}{z},\qquad \quad \frac{\partial
  F_0}{\partial a_l} = \frac{1}{2} \oint_{B_l} \log x \frac{dz}{z},
$$
acquire a standard interpretation in Seiberg-Witten theory: they relate vevs
of the Higgs field $a_l$ with the prepotential $F_0$. This completes then our
solution of the Seiberg-Witten theory from the matrix model perspective.

We also note that the matrix models which we consider are unstable due to the
unbounded potentials, as seen in figures \ref{fig-V} and
\ref{fig-V5d}. However this instability is not worse that the instability
related to the condition of fixed filling fractions, which results in unequal
level of a potential in each cut. Due to the condition of fixed filling fractions the
eigenvalues are prevented from falling into the infinite instability well in
the same way as they are prevented from tunnelling between cuts. 

%*******************************************************

\section{Relation to geometric engineering and topological string
  theory}   
\label{sec-top-string}

The purpose of this section is to give an unified geometric 
description of the 4- and 5-dimensional supersymmetric gauge 
theories by geometric engineering of type II string theory and 
M-theory on non-compact Calabi-Yau spaces. This construction 
relates the matrix models discussed in this paper and the one
in~\cite{mm-lens,lens-matrix,hal-yas}. For 4 dimensional supersymmetric 
gauge theories the subject started with the work of~\cite{KKLMV,geom-eng} 
and for the 5 dimensional cases in~\cite{Intriligator:1997pq}.
We give a short synopsis of these works focusing on the different 
geometries for the different Chern-Simons terms in the 5 dimensional
case~\cite{Intriligator:1997pq}. The main motivations are as follows. 

Physically the higher genus terms calculable in the matrix model are 
only relevant if the theory is embedded in gravity, which according 
to our present understanding is consistently possible only 
by geometric engineering of the corresponding string geometries.    

By moving in the moduli space,  which is best described as the 
complex structure moduli space of the mirror geometry, one can 
connect the local matrix model descriptions of section~\ref{sec-4d} 
and~\ref{sec-5d} to each other and to the Chern-Simons type matrix 
models that have been studied in~\cite{mm-lens,lens-matrix,hal-yas}. 

The 5-dimensional theories are labelled by different integer Chern-Simons
terms $m\in \mathbb{Z}$, which make an important difference in the 
identification with the matrix model, discussed in
section~\ref{mneq0case}. For this reason and because the absence of 
world-sheet instantons makes the theories simpler to describe, we 
discuss mainly theses cases~\cite{Intriligator:1997pq}. 

The non-compact geometries for the 5-dimensional case are the same 
as the ones used in the 4-dimensional geometric engineering after 
a limit~\cite{geom-eng}. We can therefore focus in section~\ref{sec-4d}
on the discussion of this limit, which leads to 4-dimensional 
spectral curve $H(w,z;\mu_\alpha)=0$. We explain why it does not 
depend on $m$. 

For $m=0$ and $ADE$ gauge theories $G$ there is another point in the
moduli space where one extracts the partition function of Chern-Simons
theory on the lens space $S^3/D_G$, where $D_G$ is a discrete
subgroup of $SU(2)$ acting on $S^3$. The latter have a 
corresponding $ADE$ classification and matrix model descriptions for 
these Chern-Simons theories have been suggested.
The discussion in section~\ref{4dswlimit} is necessary to specify 
the general Chern-Simons matrix model limit in~\cite{lens-matrix}.

\subsection{Geometric engineering of 5-dimensional gauge theories}    \label{ssec-geomeng}

$\mathcal{N}=1$ supersymmetric gauge theory in 5 dimensions can be
geometrically engineered from M-theory on certain non-compact 
Calabi-Yau threefolds $M$. The basic geometry of $M$ is that 
of a resolved $ADE$ singularity fibered over $\mathbb{P}^1$. 
It gives rise to a $\mathcal{N}=2$ supersymmetric gauge theory in 
five, and after a limit in four dimensions, whose gauge group $G$ is the corresponding
simply-laced group. The intersection of the ${\rm rank}(G)$ 
$\mathbb{P}^1$ curves in the fiber is the negative Cartan matrix of $G$. 
For non-simply-laced groups the fibration has to have 
non-trivial mondromy, see~\cite{Intriligator:1997pq}. Matter can be added by additional 
blow ups~\cite{geom-eng} leading to singular fibres~\cite{Intriligator:1997pq}
or by considering a higher genus base~\cite{Katz:1996ht,Klemm:1996kv}.

The precise dictionary between manifolds $M$ and the 5-dimensional gauge
theories~\cite{Intriligator:1997pq,suN-Amir,Tachikawa} contains a parameter $m\in \mathbb{Z}$, 
which labels a choice the Chern-Simons terms of the 5-dimensional
gauge theory or equivalently the triple intersections of $M$. The
choice $m=0$ corresponds to the matrix model discussed in
section~\ref{sec-5d}. Since non-compact Calabi-Yau
manifolds $M$ on which M-theory compactifications give rise to
5-dimensional gauge theories with different Chern-Simons terms we
need to describe how the differences in the non-compact 
mirror Calabi-Yau space $W$ given by  eq. (\ref{W}) are 
reflected in the matrix model. At the end we propose matrix models 
describing topological string on a wide class of non-compact toric
Calabi-Yau spaces, which correspond  to the $SU(n)$ cases.

The different 5-dimensional theories, which however lead to the same 
gauge theory in 4 dimensions, can be specified by a particular choice 
of the normalized triple intersection numbers $c_{ijk}$ of the non-compact
Calabi-Yau, which appear in the 5-dimensional theory as Chern-Simons terms
\begin{equation}
\int c_{ijk} A_i\wedge F_j\wedge F_k\ .
\label{CS}  
\end{equation}  
Here the latin indices label the vector multiplets. Note that (\ref{CS}) is
not gauge invariant, but it can be shown~\cite{Witten:1996qb} that if $M$ is 
spin with $p_1=0$, as it  is the case if $M$ is Calabi-Yau, the 5-dimensional integral over 
(\ref{CS}) shifts only by $2 \pi \mathbb{Z}$. In M-theory on a Calabi-Yau
$M$ vector fields arise as $A_i=\int_{{\cal C}_i} C^{(3)}$, i.e. by reducing
the three form field $C^{(3)}$ on a 2-cycle ${{\cal C}_i}$ in a class in
$H_2(M)$. Similarly the field strength arises as $F_i=\int_{{\cal C}_i} ({\rm d}
C)^{(4)}$, so that (\ref{CS}) is just dimensional reduction of eleven
dimensional Chern-Simons term $\int_M   C^{(3)}\wedge({\rm d} C)^{(4)}
\wedge({\rm d} C)^{(4)}$ with $c_{ijk}=\int_M \omega_i\wedge \omega_j\wedge
\omega_k= D_i\cap D_j \cap D_k$, where $\omega_i$ are in $H^2(M)$ 
with compact support on ${\cal C}_i$ and  $D_i$ are dual divisors to 
$[{\cal C}_i]$. Both definitions of the triple intersection need a 
regularization on non-compact Calabi-Yau spaces. The intersections 
are largely determined by the Cartan matrix of $G$ and 
in~\cite{Intriligator:1997pq} it was shown that the discrete 
freedom in choosing different Chern-Simons terms is a discrete choice 
$m\in \mathbb{Z}$, which parametrizes that particular triple intersection 
mentioned above. For $SU(n)$ all triple intersections will be made explicit in the
prepotential below.             

\subsection{Mirror symmetry for toric non-compact Calabi-Yau} 
Mirror symmetry implies that the data of the 5 dimensional gauge
theory as well as of the topological string on $M$ are captured by a 
Riemann surface (\ref{W}) and a meromorphic differential $\lambda$. 
One crucial point is that  spectral curve of the 5-dimensional matrix model 
discussed in the last section is identified with that Riemann surface. 
More precisely in the moduli space of the Riemann surface the 
partition function of the 5-dimensional gauge theory emerges as 
a holomorphic limit of the topological string partition function 
near the large radius point $\mu_\alpha=0$. We review below how 
the mirror curve (\ref{W}) for the geometry described in
(\ref{ssec-geomeng}) is constructed. 

The mirror manifold $W$ of the manifold $M$ is a conic bundle   
\begin{equation} 
H(w,z;\mu_\alpha) - uv =0   \label{W}
\end{equation}
branched over $\mathbb{C}^*\times \mathbb{C}^*$ at a complex family of mirror curves $H(w,z;z_\alpha)=0$,
i.e. $u,v \in \mathbb{C}$, and $w,z\in \mathbb{C}^*$. With the geometry comes a canonical 
meromorphic differential
\begin{equation}
\lambda = \log(z) \frac{dw}{w} \, .
\end{equation}
From our perspective this differential will determine the filling fractions of the matrix model. 

For the $SU(n)$ cases the well known construction of non-compact toric mirror symmetry applies 
\cite{geom-eng,Hori:2000kt}, see \cite{Haghighat:2008gw} for a short 
review in the notation we use here. The manifolds $M$ and $W$ are defined by charge 
vectors $Q^\alpha_i\in\mathbb{Z}$ of the toric group action. More
precisely 
\begin{equation} 
M=(\mathbb{C}^{k+3}-Z)/(\mathbb{C}^*)^k,
\label{quotient}
\end{equation} 
where $(\mathbb{C}^*)^k$ acts by $x_i\mapsto \lambda_\alpha^{Q_i^\alpha} x_i$,
$\alpha=1,\ldots,k$  on the complex coordinates $x_i$ of $\mathbb{C}^{k+3}$ with 
$\lambda_\alpha\in \mathbb{C}^*$. The constraint $\sum_{i=1}^{k+3} Q_i^\alpha=0$, 
$\forall \alpha$  is equivalent to $c_1(TM)=0$. $H$ is given by $H=\sum_{i=1}^{k+3} x_i$,
where\footnote{We use the same symbol $x_i$ for the different coordinates of $M$ and $W$.}
$x_i\in \mathbb{C}^*$ are homogeneous coordinates w.r.t. an additional 
$\mathbb{C}^*$-action and subject to the constraints 
\begin{equation} 
(-1)^{Q_0^\alpha} \prod_{i=1}^{k+3}x_i^{Q_i^\alpha}=\mu_\alpha\ , \qquad \qquad  \forall \alpha. 
\label{mirrormap}
\end{equation} 
Here $\mu_\alpha$, $\alpha=1,\ldots,h_{11}^{comp}$, are complex moduli of $(W,\lambda)$, which  
are dual to the complexified K\"ahler parameters of $M$, while $w$ and $z$ in (\ref{W}) denote 
the independent variables that remain after using the constraints (\ref{mirrormap}) 
as well as the additional $\mathbb{C}^*$-action on the $x_i$, mentioned above. 
$H(w,z,\mu_\alpha)=0$ is central in the following discussion, because it is identified 
with the spectral curve of the matrix model.  

\subsection{Geometric engineering of 5-dimensional $SU(n)$ gauge theories} 
\label{gaugetheorygeometries}

\begin{figure}[htb]
\begin{center}
\includegraphics[width=1\textwidth]{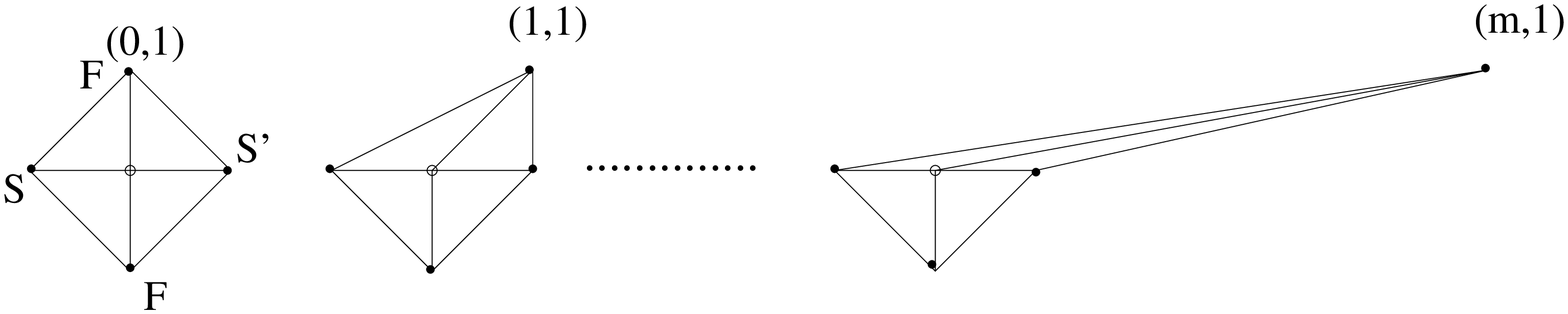}
\begin{quote}
\caption{
Toric fans for $\mathbb{F}_m$.} \label{fig-Fm}
\end{quote}
\end{center}
\end{figure}

Let us now give the dependence of the charge vectors $Q^{\alpha}_k$ 
defining $M$ and $W$ on $m$. The simplest case arises for $SU(2)$ without matter. 
It was shown in~\cite{geom-eng} that the possible 5-dimensional geometry $M$ correponds 
in this case to the anticanonical line bundle over the Hirzebruch surfaces 
$\mathbb{F}_m=\mathbb{P}({\cal O}_{\mathbb{P}^1} \oplus{\cal
  O}_{\mathbb{P}^1}(m))$, 
written here as the projectivization of line bundles over $\mathbb{P}^1$. 
The parameter $m$ in this special case is precisely the  $m\in \mathbb{Z}$ 
freedom to choose different Chern-Simons terms. 
Recall that this non-compact Calabi-Yau spaces over the Hirzebruch surfaces
$\mathbb{F}_m$ are torically described by the 3-dimensional fan spanned by 
$\{n_1,n_2,\ldots,n_5\}=\{(-1,0,1),(0,0,1)(1,0,1),(m,1,1),(1,-1,1)\}$. 
The complete fan in the plane described by first two coordinates 
with the triangulation as shown in figure \ref{fig-Fm} defines torically 
the Hirzebruch surface $\mathbb{F}_m$.

Each point  $n_i$ corresponds to a divisor class $\hat D_i=\{x_i=0\}$, where 
$x_i$ are coordinates of the (\ref{quotient}). These divisors are 
not independent, but fullfill the relations $\sum_{k=1}^r n_{k,i} \hat D_k$ 
for $r=5$ and  $i=1,\ldots, d=3$. The triangulation  specifies the 
intersection of $M$ in the following way. $\hat D_i \hat D_j=0$ if $i$ and $j$ 
are not part of one triangle. $\hat D_i \hat D_j \hat D_k=1$ if $n_i,n_j,n_k$ span a 
triangle in the triangulation. Restricted to the Hirzebruch surfaces one has 
$\sum_{k=1}^r n_{k,i} D_k$  for $r=5$ and  $i=1,\ldots, d=2$, 
$D_i\cap D_j=1$, $i,j>1$ if $n_i,n_j$ are on a line\footnote{We use hatted
characters for non-compact divisors of $M$ that restrict to compact divisors
on $\mathbb{F}_m$.}. There are two independent classes in $\mathbb{F}_m$, 
$F=D_4=D_5$ and $S=D_1=S'-mF$ with $S'=D_3$. 
With the above information and the ring structure of the intersections 
it is easy to calculate all toric intersections. For the Hirzebruch surface 
$FS=1$, $F^2=0$, $S'S=0$ and $S^2=-2$. For $M$ there is now a 2-dimensional complex
dimensional compact divisor $H=D_1$, which represents the Hirzebruch surface itself. 
$H^3=4$, $H\hat S^2=-m$, $H \hat F^2=0$  and $H\hat F\hat S=1$, where 
$\hat S$ and $\hat S$ are non-compact divisors of $M$.
Note that the triple intersections among the non-compact divisors 
corresponding to  points on the boundary of the fan are not well defined.         

The charge vectors are coefficients of linear relations among the points
$\sum_i Q^\alpha_i n_i=0$. 
The one which do actually span the Mori 
cone are calculated using the secondary polytop to be 
$Q^1=(1,-2,1,0,0)$ and  $Q^2=(0,m-2,-m,1,1)$. They correspond to curve classes 
and the intersection number of the $\alpha$'th curve with the divisor $D_i$ is
encoded in $Q^\alpha_i$. The K\"ahler cone is dual to the Mori cone and it is
easy to see that it is spanned by $F$ and $H=S+(m+1)F$. 

These geometries are easily generalized to $SU(n)$. To distinguish $n$ odd and
$n$ even we define $\Delta(n)=\frac{1}{2}(1-(-1)^n)$. The    
points of the 3-dimensional fan  and the Mori cone are given, 
with $n_l=\frac{\Delta(n)-n}{2}$ and $n_u=\frac{n+\Delta(n)}{2}$, as 
$$  %\begin{equation}
\begin{array}{c|cccccccccccccc}
                  n_i& Q^1 & Q^2 & Q^3 & \ldots   &  & & &      & & & &Q^n\\  \hline       
\left(n_l ,0,1\right)&1 &0 &0   &\ldots &0 &0&0&\ldots&0&0&0&0\\[-2mm]
\left(n_l+1, 0,1\right)&-2&1 &0   &\ldots &0 &0&0&\ldots&0&0&0&0\\[-2mm]
\left(n_l +2,0,1\right)&1 &-2&1   &\ldots &0 &0&0&\ldots&0&0&0&0\\[-2mm]
                          \vdots&\vdots&\vdots& \vdots &\ldots  &\vdots&\vdots&\vdots&\ldots &\vdots&\vdots&\vdots&\vdots\\[-2mm]
\left(0,0,1\right)              &0 &0 &0   &\ldots &1&-2&1&\ldots&0&0&0&(m-2+\Delta(n))\\ [-2mm]
\left(1,0,1\right)              &0 &0 &0   &\ldots &0&1&-2&\ldots&0&0&0&-m-\Delta(n)\\ [-2mm]
                          \vdots&\vdots&\vdots& \vdots &\ldots  &\vdots&\vdots&\vdots&\ldots &\vdots&\vdots&\vdots&\vdots\\[-2mm]
\left(n_u-2,0,1\right)&0 &0&0   &\ldots &0 &0&0&\ldots&1&-2&1&0\\[-2mm]
\left(n_u-1,0,1\right)&0 &0&0   &\ldots &0 &0&0&\ldots&0& 1&-2&0\\[-2mm]
\left(n_u,0,1\right)&0 &0&0   &\ldots &0 &0&0&\ldots&0& 0&1&0\\[-2mm]
\left(m,1,1\right)&0 &0&0   &\ldots &0 &0&0&\ldots&0& 0&0&1\\[-2mm]
\left(\Delta(n),-1,1\right)&0 &0&0   &\ldots &0 &0&0&\ldots&0& 0&0&1\\
\end{array} 
%\label{mori}
$$ %\end{equation} 

\begin{figure}[htb]
\begin{center}
\includegraphics[width=1\textwidth]{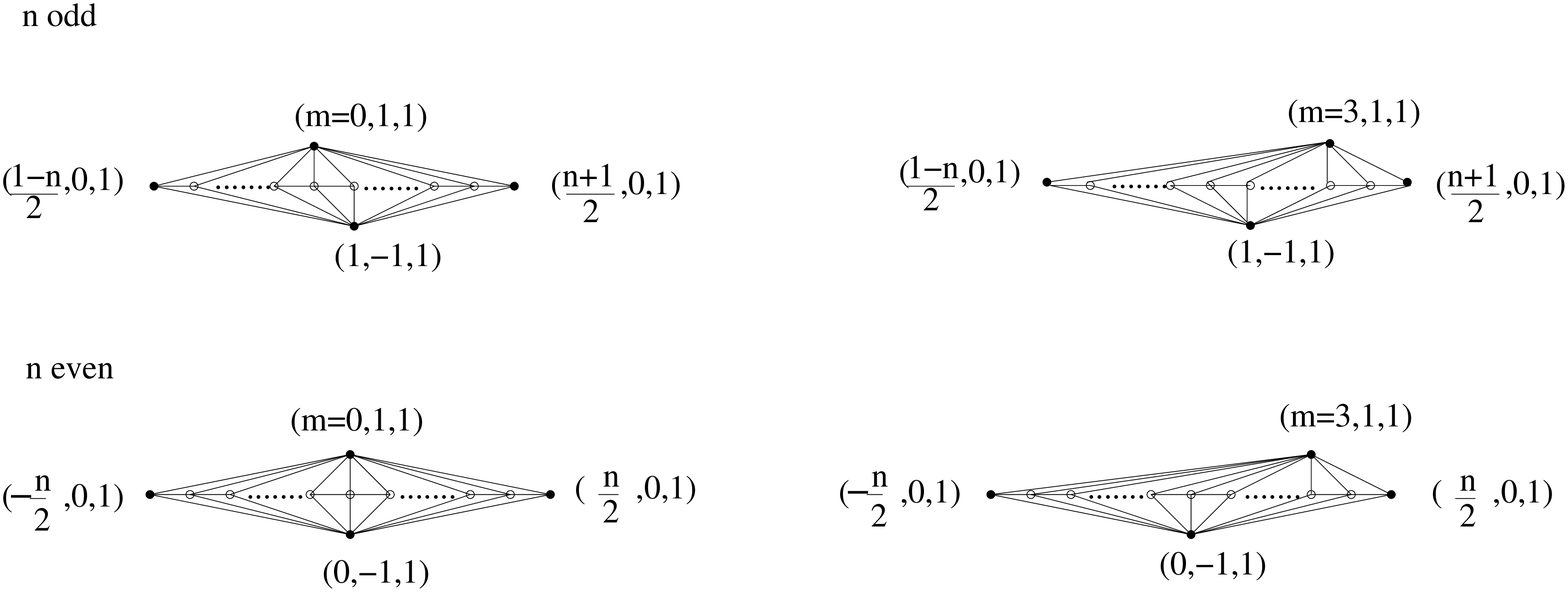}
\begin{quote}
\caption{
Toric fans for the non-compact toric Calabi-Yau corresponding to $SU(n)$
theory with the Chern-Simons term parameterized by $m$.} \label{fig-fans}
\end{quote}
\end{center}
\end{figure}
 
One can see from the pictures that the inner points in the diagram correspond
to toric compact divisors of the type $\mathbb{F}_k$  in the non-compact
threefold, which can be easily seen to be Hirzebruch surfaces. 
We denote these divisors $H_i$, $i=1,\ldots,N-1$. From the left to the
right $H_i=\mathbb{F}_{|m+2 i-n|}$, $i=1,\ldots,n-1$. With the description for
the calculation of the intersection of the $H_i$ it is easy to confirm~~\cite{Intriligator:1997pq} 
that the cubic prepotential, which is not corrected by intantons for 5-dimensional theories is
$F=\frac{1}{2} \sum_{i<j} |a_i-a_j|^3+ m \sum_{i} a_i^3=\sum_{i_1,i_2,i_3}
H_{i_1} H_{i_2} H_{i_3} \prod_{r=1}^3 (a_{i_r+1}+\ldots+a_{i_n})$.

The standard parametrization of mirror symmetry using the Mori cone in
(\ref{mirrormap}) implies that $\mu_\alpha=0$ is a point of maximal unipotent monodromy
and that near this point  $t_\alpha\sim \frac{1}{2 \pi i} \log(\mu_\alpha)$, with $t_\alpha$ the 
compexified volumes of the curves corresponding to the Mori cone. For
$\alpha=1,\ldots,n-1$ these curves are the fibres of the Hirzebruch surfaces. 
Geometrically their size is proportional to the mass of the $W$ bosons in
string units $M_{string}$. It is given by the simple roots, 
i.e. $t_\alpha=a_\alpha-a_{\alpha+1}$, $\alpha=1,\ldots,n-1$. The parameter $t_b:=t_n$ 
corresponds to the complexified size of the base $\mathbb{P}^1$. The size
appears as imaginary part while the $B$-field integral over the 
curve is the real part.
         
\subsection{4-dimensional Seiberg-Witten limit} 
\label{4dswlimit}
We come now to the engineering of the 4-dimensional gauge theory. The aim is to derive
from the 5-dimensional spectral curve $H(w,z;\mu_\alpha)=0$ in a limit 
the 4-dimensional Seiberg-Witten (\ref{sw}) curve and show that $\lambda$ reduces to
$dS$. Moreover in the 4-dimensional limit the differences given 
by the parameter $m$ should vanish.  While the 5-dimensional gauge theory discussed above is 
recovered in the large radius limit, the construction of the
4-dimensional gauge theory requires a double scaling limit 
in a region of the moduli space, called Seiberg-Witten point, 
where world-sheet instantons dominate. In fact it has been 
made explicit in \cite{geom-eng} how the individual gauge theory 
instantons contributions of Seiberg-Witten gauge theory  are encoded in a 
resummation of the world-sheet intantons.                

Non-abelian gauge groups in string theory have been first 
understood in the heterotic construction. For finding 
the Seiberg-Witten point it is instructive to recall the duality 
between the heterotic and type II strings. It states~\cite{KKLMV}
that the heterotic dilaton $S$ is geometrized and identified 
with $\frac{1}{4 \pi i}t_b$, i.e. the size of the base of 
the fibration discussed in \ref{gaugetheorygeometries}. In the heterotic
string one has ${\rm Re}(S)=\frac{1}{g^2}$ so that 
${\rm Im}(t_b)=\frac{4 \pi} {g^2}$. The asymptotic free $g\rightarrow 0$ 
region of Seiberg-Witten must be at large volume of the base $\mathbb{P}^1$ 
and hence near $\mu_b:=\mu_n=0$. However a double scaling limit, parametrized 
below by $\epsilon\rightarrow 0$ must be taken to keep the mass of the $W$ 
bosons finite, while taking the string scale to infinity. 
The mass of the $W$ bosons is the only scale and the one-loop running 
of the gauge coupling in $\mathcal{N}=2$ gauge theory  
implies $\frac{1}{g^2}\sim\log\left(\frac{M_W}{\Lambda}\right)$. Since
$W$-boson comes from branes wrapping the fibers one has $t_b\sim C \log(t_f)$, 
where $t_f$ refers to generic volumes of the fiber.
The constant $C$ can be determined from the dependence
of the instanton contribution with instanton number $k$ namely 
$\exp\left(-\frac{k}{g^2}\right)\sim \left(\frac{1}{\hat t_f^{2n}}\right)^k$, 
where $\hat t_f$ refers volumes of the fibers, i.e. masses of the $W$ bosons, 
in gauge theory units. Putting the above facts together, the double scaling 
limit is $\exp(t_b)=(\epsilon \Lambda)^{2n}$ and 
$t_f=\epsilon \hat t_f$. In the B-model geometry, i.e. the manifold $W$, $t_f$ is 
mapped to period integrals over $S^1$, which come from periods over $S^3$ 
in the compact geometry. That is the Seiberg-Witten curve is near 
point where the divisor $\mu_n=0$ intersects $n-1$ conifold divisors
in the moduli space. The double scaling limit can also be viewed as resolving the 
point of tangency between the divisors of this intersection 
point, to obtain normal crossing divisors.

Let us see for the $SU(2)$ cases how (\ref{sw}) and $dS$ is recovered 
in the limit from the $5d$ curve. According to the table in section \ref{gaugetheorygeometries} and the  
description how to obtain $H(w,z; \mu_\alpha)$ from (\ref{mirrormap}) we get   
\begin{equation} 
H(w,z; \mu_\alpha)=w+\frac{\mu_f z^{2-m}}{w}+1+ z + \mu_b z^2=0
\end{equation}     
as the defining equation of the 5-dimensional curve. Here we identified in $H=\sum_{i=1}^5 x_i$,
$w:=x_5$ and $z:=x_2$. We set $x_3=1$ by the $\mathbb{C}^*$-action
and eliminated $x_4$ and $x_1$ by (\ref{mirrormap}). Now we calculate the
conifold discriminant and find that it behaves like $\Delta=(1-4 \mu_f)^2 
+ {\cal O}(\mu_b)$, where all the $m$-dependence is in higher order in 
$\mu_b$. $M_W^2$ is proportional to $\Delta$ which and the scaling argument implies 
that the finite gauge theory coordinate $u\sim \hat t_f^2 $ should appear as 
\begin{equation}
\begin{array}{rl}  
\eta_1&=(1-4 \mu_f)=\epsilon^2 u, \\ [2 mm]  
\eta_2&=\displaystyle{\frac{\sqrt{\mu_b}}{(1-4 \mu_f)}=\frac{\Lambda^2 e^{-S_0/2}}{u}},
\end{array}
\label{blowupvariables} 
\end{equation} 
here $e^{-S_0/2}$ is an a priori undetermined constant 
and (\ref{blowupvariables}) are just the blow up variables 
to resolve the tangency between $\mu_b=0$ and $\Delta=0$.
An important point is that $z$ is a $\mathbb{C}^*$ variable in 
$H(w,z;\mu_\alpha)=0$, but in $(\ref{sw})$ $x$ is a $\mathbb{C}$ variable, which  
is clearly reflected by the $\log(z)$ dependence of $\lambda $ as opposed to 
the $x$ dependence of $dS$. So we can think of (\ref{sw}) as a compactification 
of $H=0$ in a special limit of its parameter space. To match the the 
differentials we must identify in leading order $z=1-\epsilon \tilde
x$. Rescaling $\tilde x = \frac{\hat x}{\epsilon \sqrt{af}}$ and 
eliminating the $\tilde x$ term by $\tilde x = x + \frac{1}{2}
\left(\frac{1}{\sqrt{\mu_f}} + 2 \sqrt{\mu_f}\right)$ one gets independent of
$m$ in leading order $w+\frac{\sqrt{\mu_b}}{w}= -x^2 - \left(1 - \frac{1}{4 \mu_f}\right)$. 
Rescaling further $w\rightarrow w \sqrt{\mu_b}$ and $x\rightarrow \frac{i
(\mu_b)^\frac{1}{4}}{\Lambda} x $ and using (\ref{blowupvariables}) we get (\ref{sw}).

\subsection{Matrix model description of the $m\neq 0$ cases}
\label{mneq0case}

For $m\neq 0$ the microscopic matrix model can be obtained starting with the 
modified Nekrasov partition function~\cite{Tachikawa}, which is written as in
(\ref{Nekrasov-PF-5d}) with a linear modification and quadratic modification in $\lambda$ by the
second Casimir operator
\begin{equation} 
\kappa_\lambda=\sum_{i} \lambda_i(\lambda_i+1-2i)
\end{equation} 
and a linear term in $\lambda$ as 
\be
Z^{SU(n)}_{5d}  =  \sum_{\vec{\lambda}=(\lambda^{(1)},\ldots,\lambda^{(n)})} (\beta\Lambda)^{2n 
|\vec{\lambda}|} e^{- m \beta \sum_{l} |\lambda^{(l)}| a_l+\hbar \kappa_{\lambda^{(l)}}}  Z_{\beta,\vec{\lambda}}.
\label{5d-m=!=0}\ee
If is straightforward to check~\cite{Tachikawa}, that this expression is
reproduced by the applications of the topological vertex~\cite{vertex} to the 
geometries discussed in section~\ref{gaugetheorygeometries}. 

Likewise it is simple to see, that if we apply the same steps as in 
section \ref{sec-5d} to  $ Z_{\beta,\vec{\lambda}}$ we get (\ref{V-5d}), 
except that we cannot view the whole expression, which follows 
from (\ref{5d-m=!=0}) as a matrix integral, because of the different 
linear coupling (distinguished by $a_l$) to $n$ sets of eigenvalues in the 
potential. These linear terms are the only relevant change in the 
potential, since the shifts in the definition (\ref{hk-5d}) give just rise 
to an overall constant and  moreover the quadratic terms from the second
Casimir operator will vanish in the $\hbar\rightarrow 0$ limit.    

In order to convert this to matrix model 
one has to introduce $n$ matrices. The procedure is similar 
as in~\cite{lens-matrix}. A linear term of the form $a_l\sum_{i} u_i^{(l)}$ in
a quadratic potential, which is integrated against the deformed measure 
$\Delta_q(u^{(l)})$  (\ref{Van-sinh}) can be absorbed into an 
interaction between the different groups of eigenvalues
$\Delta_q(u^{(l)},u^{(k)})=\prod_{i,j} 2{\rm sinh}
\left(\frac{u_i^{(l)} -u_j^{(k)}}{2}\right)$ in the measure. 
This holds for the potential  (\ref{SW-V5d}). After viewing 
the $n$ groups of $u^{(l)}$, $l=1,\ldots,n$ as the eigenvalues of $n$ matrices 
the interacting factors in the measure can be exponentiated to give a relatively 
complicated interaction term between the matrices~\cite{lens-matrix}. 
The matrix models for $m\neq 0$ describe, e.g. for $Su(2)$ and $m=1$ the
anticanonical bundle over $\mathbb{F}_1$, which by a blow down becomes 
$\mathbb{P}^2$. So it seems possible to write down a microscopic multimatrix model,
which reproduces in a limit the simplest non-compact toric Calabi-Yau 
${\cal O}(-3)\rightarrow \mathbb{P}^2$.

%*******************************************************
%*******************************************************

\section{Relations to other matrix models}   \label{sec-models}

In this section we show that various well-known and seemingly unrelated matrix
models arise in various limits of the 5-dimensional matrix model derived in section
\ref{sec-5d}. These interesting cases include Chern-Simons matrix models for
lens spaces, as well as matrix models for line bundles over $\mathbb{P}^1$. Moreover,
in the 4-dimensional limit the 5-dimensional model reduces to the matrix model from section
 \ref{sec-4d}, which, as we discuss below, can be regarded as a generalization
 of the Eguchi-Yang matrix model (which itself corresponds just to the $n=1$ case).
In a sense, in this way we exhibit a part of a vast and unified matrix model landscape, 
whose various limits were introduced and considered in the past in an unrelated and independent fashion.
We discuss these limiting cases in the following subsections.

\subsection{Chern-Simons matrix model for lens spaces}
 
The orbifold limit in the $m=0$ case is at the point in the moduli space where
all $t_i$, $i=1,\ldots,n$, are small. In the complex structure variables this 
is near the vanishing of $w_i=1-\frac{z_i}{z_{i+1}}$, $i=1,\ldots, n-1$ and
$w_n=\frac{1}{\sqrt{z_{n-1}} \left(1-\frac{z_n}{z_{n-1}}\right)}$. 
Here is another matrix model known describing this topological string theory 
in the orbifold point \cite{lens-matrix}. Its spectral curve was analyzed 
in \cite{hal-yas,hal-ok-yas} and claimed to be the curve $H(x,y;z)=0$ 
obtained from (\ref{mirrormap}) and the table in section \ref{gaugetheorygeometries}, as described above.
 
The matrix model presented in \cite{lens-matrix} looks almost the same as ours -- the only
difference is that its potential contains only the quadratic term. It does not contain
neither the linear term in $t$, nor the dilogrithms. The absence of the linear term is
easy to explain, as $t=0$ should in principle correspond to the orbifold limit. 
We also note that the dilogarithms are suppressed when their arguments have large negative values. 
It would be interesting to understand the relation between a continuation from the large radius 
to the orbifold point in more detail. We also note that slightly more general models for orbifold points 
corresponding to arbitrary lens spaces $L(p,q)$ we recently found in \cite{matrix-Lpq}. They also have 
just the quadratic potential and the deformed measure, so belong to the same family of models as those described above.

\subsection{Matrix models for line bundles over $\mathbb{P}^1$}   \label{ssec-bundleP1}

In \cite{CGMPS} topological strings for Calabi-Yau manifolds of the form
$$
X_p = \mathcal{O}(p-2)\oplus\mathcal{O}(-p) \to \mathbb{P}^1
$$
were considered. Partition functions for such theories can be computed
using the topological vertex \cite{vertex}
$$
Z_{X_p} = \sum_{\lambda} C_{\lambda\bullet\bullet}C_{\lambda^t\bullet\bullet} q^{(p-1)\kappa_{\lambda}
  /2}(-1)^{|\lambda|} e^{-t|\lambda|},
$$
where the sum is performed over all two-dimensional partitions $\lambda$,
$\kappa_{\lambda}=\sum_i \lambda_i(\lambda_i -2i +1)$ is the
second Casimir, $q=e^{-g_s}$, and $t$ is the K{\"ahler} parameter. The vertex
amplitude in this case $C_{R\bullet\bullet} = P_q(\lambda)^{1/2}$ is a square root
of the deformed Plancherel measure (\ref{planch-q}), so it leads to the same
partition function as considered in \cite{eynard-planch}. The case $p=1$ can be considered as the
5-dimensional $U(1)$ theory, which is a special case $n=1$ of our general result. It
was shown in \cite{CGMPS} that the above partition function can be written as
a one-cut matrix model with a deformed measure and the potential
$$
V(u) = (t-(p-1))u + \frac{p}{2}u^2 + 2\,\Li_2(e^{u}).
$$
For $p=1$ this is indeed the potential which we obtain in $n=1$ case. For
arbitrary $p$ only the linear and quadratic terms are modified (the latter due
to the Casimir $\kappa_{\lambda}$ in $Z_{X_p}$).

\subsection{Eguchi-Yang matrix model}

We also note that in the limit of the 4-dimensional theory and for $n=1$ our matrix model with
the potential (\ref{SW-V4d}) is closely related to the $\mathbb{CP}^1$ matrix model introduced by Eguchi 
and Yang
\bea
Z^{EY} & = & \int \mathcal{D}M e^{-2N \Tr \,V^{EY}(M)}, \nonumber \\
V^{EY} & = & (-M+M\log M) - \sum_{i=1}^{\infty} t_{i-1,Q} M^i -
\sum_{i=1}^{\infty} t_{i,P} M^i(\log M - \sum_{j=1}^i \frac{1}{j}).  \nonumber
\eea
This matrix model was introduced in \cite{E-Y,E-H-Y}, where it was also shown
that its partition function $Z^{EY}$ reproduces the partition function of the
topological $\mathbb{CP}^1$ model considered in \cite{dw-cp1}, i.e. the topological
sigma-model coupled to the two-dimensional gravity. This model has two
observables denoted $P\equiv\sigma_0(P)$ and $Q\equiv\sigma_0(Q)$, which
correspond respectively to the identity and the K{\"a}hler class of
$\mathbb{P}^1$. In addition, two infinite series of descendants $\sigma_i(P)$
and $\sigma_i(Q)$, for $i=1,2,\ldots$, arise in this theory upon coupling
to gravity. Coupling constants corresponding to these descendants are denoted
by $t_{i,P}$ and $t_{i,Q}$, and they appear in the Eguchi-Model as coefficients
  in two infinite series in the potential. 

We immediately see that a special case of the Eguchi-Yang model corresponding
to $t_{i,P}=t_{i,Q}=0$ for $i \geq 1$ is identical with the $n=1$ case of our matrix model
(\ref{SW-V4d}), with our $t$ identified with the coupling $t_{0,Q}$ in
$V^{EY}$. From physical point of view the vanishing of all coupling
constants corresponds to replacing a matrix $M$ by a chiral superfield
\cite{cet-va}. From the model with $n=1$ and vanishing coupling constants one
can derive the limit shape of partitions given by the arcsin-law
\cite{eynard-planch,Mar-Nek}. Our matrix 
model for Seiberg-Witten theory (\ref{SW-V4d}) is therefore a generalization of the
Eguchi-Yang matrix model to arbitrary $n$. More generally, setting all
$t_{i,P}=0$ but taking arbitrary $t_{i,Q}$ is identical with the matrix model
considered in \cite{eynard-planch}, with arbitrary Casimirs taken into account. 
We note that some subtleties, adventages and drawbacks of the comparison 
of the sums over partitions arising from gauge theory instanton calculus, 
to the matrix model of the Eguchi-Yang type,
are also discussed in \cite{Mar-Nek}.

\section{Conclusions and discussion}  \label{sec-disc}

In this paper we provided a formulation of Seiberg-Witten theory in terms of
matrix models. For 4-dimensional theories we derived 1-matrix models with a
standard measure containing the Vandermonde determinant, while the 1-matrix models 
for 5-dimensional theories have a deformed measure. We also extended 
this approach to a matrix model formulation of topological string theory on 
Calabi-Yau manifolds which are related to Seiberg-Witten theories 
by geometric engineering. In general this involves multi-matrix models. 

The 1-matrix models which we propose were derived in two steps. Firstly, we
introduced in the Nekrasov partition function an auxiliary parameter $N$, 
and rewrote this partition function as a matrix integral over matrices of size $N$. 
The matrix models obtained this way are rather non-standard 1-matrix models,
with potentials which contain such special functions as logarithms of $\G$
functions. Then, the partition functions of the Seiberg-Witten theory were
recovered in the 't Hooft limit. In this limit various simplifications
take place and we end up with matrix potentials containing logarithmic and
dilogarithmic terms. 

The derivation of these matrix models is just an initial step, which
should be followed by a detailed analysis of their properties. We 
use the Migdal-Muskhelishvili formula for the resolvent to check that the 1-matrix 
models for the 5-dimensional theories yield the expected spectral 
curves with the correct differential $\lambda$ defining the filling 
fractions. The 4-dimensional spectral curves and the corresponding 
differential $dS$ are reproduced by in the gauge theory limit of geometric 
engineering. It could have also been derived with the Migdal-Muskhelishvili 
formalism for the resolvent in 4-dimensional theory directly. 

The general framework for solving matrix models based just on 
the underlying spectral curve and the differential, which
was developed by Eynard and Orantin~\cite{eyn-or,eynard} and checked 
in very similar situations in~\cite{marino-eo,remodel}, should give the higher
genus, as well as open amplitudes in all our cases\footnote{Some checks in 
4-dimensional Seiberg-Witten theory have been made in~\cite{HKII}.}. The holomorphic 
anomaly equations and the gap condition applies likewise 
to the 4-dimensional cases~\cite{HK} and to the 5-dimensional models~\cite{Haghighat:2008gw}. 
It provides the most efficient way to calculate the closed higher genus amplitudes.  

Certainly a perturbative expansion of the matrix models discussed here is
possible. Of course we hope that the knowledge of the microscopic matrix model 
action will provide a route towards better understanding non-perturbative
effects in Seiberg-Witten theory and topological strings. Techniques
of analyzing non-perturbative effects in matrix models, being currently
developed by~\cite{marino-nonpert1,marino-nonpert2}, require the knowledge of
the potential, which we provide in this paper. They would also be of great 
interest from the point of view of black hole physics and OSV
conjecture \cite{OSV}, and complementary to the proposition 
discussed in \cite{qYM-AOSV,qYM-AJS}.

An important aspect of the matrix models derived here is their relation to
other families of matrix models. Firstly, matrix models for the $SU(n)$
4-dimensional Seiberg-Witten theory can be regarded as a generalization of the the Eguchi-Yang 
matrix model (which corresponds to $n=1$). Higher Casimir operators, which
we did not analyse in this paper, but in the context of $n=1$ model they were
taken into account into \cite{eynard-planch,Mar-Nek}, would correspond to
gravitational descendants in the Eguchi-Yang model. In the same spirit, matrix models
for 5-dimensional $SU(n)$ theory derived here are generalizations of the matrix model
for bundles over $\mathbb{P}^1$ considered in \cite{CGMPS} (which also
corresponds to $n=1$). It is also reassuring that the models which we derived
contain no other terms than logarithms and dilogarithms, which also appeared
in \cite{CGMPS}. While our models, as well as those in \cite{CGMPS},
correspond to the large radius point in the moduli space, it is also possible
to consider other regions of the moduli space. In particular, in the limit
where dilogarithms in the potential for the 5-dimensional theory are
suppressed, we recovered Chern-Simons matrix models for lens spaces which
correspond to orbifold points in the moduli space
\cite{lens-matrix,hal-yas,hal-ok-yas}. The similarity of the microscopic
description at the large radius point and at the orbifold point (just the 
suppression of dilogarithmic terms) is surprising in view  of 
highly involved transformations of partition functions between 
these points.

The above results indicate a possible unified microscopic description of
the multitude of matrix models, which describe various regions in the moduli 
space. Similar ideas of the unification of matrix models to some overall
matrix-model-M-theory (in analogy to string theory) were advocated 
in~\cite{M-matrix-1,M-matrix-2}.

Our results should also provide a direct link between fermionic formulations of
matrix models, gauge theories and topological strings. The fermionic viewpoint 
on matrix models, intimately related to conformal field theory techniques, was
originally advanced by Kostov \cite{cft-matrix}. Recently the (matrix model) formalism of
Eynard and Orantin was also rephrased in a fermionic language 
\cite{dv-eo-ks}. On the other hand, topological strings and gauge theories can
be formulated in terms of two-dimensional chiral fermions too
\cite{Nek-Ok,vertex,dhsv,dhs,phd,deform}. Such fermions live on curves which, in the present
context, are simultaneously matrix model spectral curves and Seiberg-Witten
curves. This convinces that all those fermions have a common origin. Moreover,
in \cite{dhs} a formalism was presented which relates both matrix models and 
Seiberg-Witten theory to $\mathcal{D}$-modules. Finding how matrix models
presented in this paper fit into this formalism would certainly be of
interest. 

Let us also stress that the idea of expressing gauge theory quantities in terms of matrix integrals has been 
explored to much extent in the context of $\mathcal{N}=1$ theories within the so-called Dijkgraaf-Vafa formalism 
\cite{DV-0206,pert-window}. $\mathcal{N}=1$ theories can be treated as deformations of $\mathcal{N}=2$ theories,
and it was also shown that certain scaling of Dijkgraaf-Vafa matrix models leads to the $\mathcal{N}=2$ answers 
\cite{pert-window,N-1-2-fluxes}. The difference with the case considered in this paper is that now we get the 
answer which gives $\mathcal{N}=2$ results without a need for any
rescaling. Some other matrix models or related ideas corresponding to gauge theories were also
proposed in \cite{hollowood,wijnholt,ferrari,so,tai}. It would be interesting to relate these points of view.

%*******************************************************
%*******************************************************

%\newpage
\bigskip

\bigskip

\centerline{\Large{\bf Acknowledgments}}

\bigskip

We would like to thank Rainald Flume for many comments, suggestions
and collaboration in the initial stages of this project. We also appreciate
useful suggestions and discussions with Vincent Bouchard, Marcos Mari{\~n}o,
Andrei Marshakov,  Andrei Mironov, Nikita Nekrasov and Cumrun Vafa. 
P.S. would like to thank the organizers of the 6$^{th}$ Simons
Workshop in Mathematics and Physics in Stony Brook, were parts of this project
were done, for hospitality and inspiring atmosphere. The research of P.S. was
supported by the Humboldt Fellowship.

\bigskip

%*******************************************************
%*******************************************************

\newpage

\appendix

\bigskip

\centerline{\Large{\bf Appendix - elliptic integrals}}

\bigskip

To derive the spectral curve for 5-dimensional $SU(2)$ theory we need to
perform the following integral
$$
\int^v \frac{v^{-1} (z-v)^{-1}\, dv}{\sqrt{(v-a)(v-b)(v-c)(v-d)}}
= g_1 F(y(v),k) + g_2 \Pi(y(v),\delta(0),k) + g_3\Pi(y(v),\delta(z),k),
$$
which is expressed as a combination of the elliptic functions, with the
following notation. The coefficients in this combination are equal
\be
g_1 = \frac{2}{a(a-z)\sqrt{(b-c)(a-d)}}, \qquad  \qquad
g_2 = \frac{2(b-a)}{z a b \sqrt{(b-c)(a-d)}}, \label{g123}
\ee
$$
g_3 = \frac{2(b-a)}{z (z-a)(z-b)\sqrt{(b-c)(a-d)}},
$$
while the arguments of $F$ and $\Pi$ are given as
\be
y(v) = \frac{(a-d)(b-v)}{(b-d)(a-v)},\qquad \delta(z) = \frac{1}{y(z)},\qquad
k^2 = \frac{1}{y(c)} = \frac{(b-d)(a-c)}{(a-d)(b-c)}.   \label{ydeltak2}
\ee

The elliptic functions $F$ and $\Pi$, of the first and the third kind
respectively, are
\be
F(y,k) =  \int_0^y \frac{dt}{\sqrt{(1-t^2)(1-k^2 t^2)}}, \qquad 
\Pi(y,\delta,k) = \int_0^y \frac{dt}{(1-\delta t^2)\sqrt{(1-t^2)(1-k^2
    t^2)}}. 
\ee
The complete elliptic integrals are defined as $K(k) \equiv F(1,k)$ and
$\Pi(\delta,k) \equiv \Pi(1,\delta,k)$. Of particular importance in our
computation are the following addition formulas \cite{byrd}
\bea
F(\sin \vartheta,k) \pm F(\sin \beta,k) & = & F(\sin \varphi,k), \label{addF} \\
\Pi(\sin \vartheta,\delta,k) \pm \Pi(\sin \beta,\delta,k) & = & \Pi(\sin
\varphi,\delta,k) \pm \sqrt{\frac{\delta}{(\delta-1)(\delta-k^2)}}\, \textrm{artanh}\, \Theta,   \label{addPi} 
\eea
where
\be
\Theta = \frac{\sin\vartheta\,
\sin\beta\,\sin\varphi\,\sqrt{\delta(\delta-1)(\delta-k^2)}}{1-\delta\sin^2\varphi +
\delta\sin\vartheta\,\sin\beta\,\cos\varphi\,\sqrt{1-k^2\sin^2\varphi}},   \label{kat-Theta}
\ee
and the angle $\varphi$ is determined as
\be
\cos \varphi = \frac{\cos\vartheta\,\cos\beta \mp \sin\vartheta\sin\beta\,\sqrt{(1-k^2\sin^2\vartheta)(1-k^2\sin^2\beta)}}
{1-k^2\sin^2\vartheta\,\sin^2\beta}.   \label{kat-phi}
\ee

%*******************************************************************
%*******************************************************************

\end{document}